\documentclass[numberedappendix,apj,twocolumn]{emulateapj}

\newcommand{\bFilter}{$B_{435}$}
\newcommand{\vFilter}{$V_{606}$}
\newcommand{\iFilter}{$i_{775}$}
\newcommand{\zFilter}{$z_{850}$}
\newcommand{\yFilter}{$Y_{105}$}
\newcommand{\jhFilter}{$JH_{140}$}

\newcommand{\fluxunit}{erg\,s$^{-1}$\,cm$^{-2}$}

\newcommand{\iwFilter}{$I_{814}$}

\newcommand{\jFilter}{$J_{125}$}

\newcommand{\hFilter}{$H_{160}$}

\newcommand{\msol}{$M_\odot$}

\newcommand{\Sp}{{\it Spitzer\/}}

\shorttitle{Bright $z\sim9-10$ Galaxy Candidates in GOODS-North}
\shortauthors{Oesch et al.}

\begin{document}

\title{The Most Luminous $z\sim9-10$ Galaxy Candidates yet Found:  The Luminosity Function, Cosmic Star-Formation Rate, and the First Mass Density Estimate at 500 Myr\altaffilmark{1}
}

\altaffiltext{1}{Based on data obtained with the \textit{Hubble Space Telescope} operated by AURA, Inc. for NASA under contract NAS5-26555. Based on observations with the Spitzer Space Telescope, which is operated by the Jet Propulsion Laboratory, California Institute of Technology under NASA contract 1407.}

\author{P. A. Oesch\altaffilmark{2,3,\dag},
R. J. Bouwens\altaffilmark{4}, 
G. D. Illingworth\altaffilmark{2}, 
I. Labb\'{e}\altaffilmark{4}, 
R. Smit\altaffilmark{4},
P. G. van Dokkum\altaffilmark{3}, \\
I. Momcheva\altaffilmark{3},
M. L. N. Ashby\altaffilmark{5}, 
G. G. Fazio\altaffilmark{5}, 
J.-S. Huang\altaffilmark{5}, 
S. P. Willner\altaffilmark{5},
V. Gonzalez\altaffilmark{6},\\
D. Magee\altaffilmark{2},
M. Trenti\altaffilmark{7},
G. B. Brammer\altaffilmark{8},
R. E. Skelton\altaffilmark{9},
L. R. Spitler\altaffilmark{10,11}
}

\altaffiltext{2}{UCO/Lick Observatory, University of California, Santa Cruz, 1156 High St, Santa Cruz, CA 95064; poesch@ucolick.org}
\altaffiltext{3}{Yale Center for Astronomy and Astrophysics, Yale University, New Haven, CT 06520}
\altaffiltext{4}{Leiden Observatory, Leiden University, NL-2300 RA Leiden, Netherlands}
\altaffiltext{5}{Harvard-Smithsonian Center for Astrophysics, Cambridge, MA, USA}
\altaffiltext{6}{University of California, Riverside, 900 University Ave, Riverside, CA 92507, USA}
\altaffiltext{7}{Institute of Astronomy and Kavli Institute for Cosmology, University of Cambridge, Cambridge CB3 0HA, UK }
\altaffiltext{8}{Space Telescope Science Institute, 3700 San Martin Drive, Baltimore, MD 21218, USA}
\altaffiltext{9}{South African Astronomical Observatory, P.O. Box 9, Observatory 7935, South Africa} 
\altaffiltext{10}{Department of Physics and Astronomy, Faculty of Sciences, Macquarie University, Sydney, NSW 2109, Australia} 
\altaffiltext{11}{Australian Astronomical Observatory, PO Box 915, North Ryde, NSW 1670, Australia}
\altaffiltext{\dag}{Hubble Fellow, YCAA Fellow}

\begin{abstract}

We present the discovery of four surprisingly bright ($H_{160} \sim 26 - 27$ mag AB) galaxy candidates at $z\sim9-10$ in the complete HST CANDELS WFC3/IR GOODS-N imaging data,
doubling the number of $z\sim10$ galaxy candidates that are known, just $\sim$500 Myr after the
Big Bang. Two similarly bright sources are also detected in a systematic re-analysis of the GOODS-S data set.
Three of the four galaxies in GOODS-N are significantly detected at $4.5-6.2\sigma$ in the very deep $Spitzer$/IRAC 4.5\,$\mu$m data, as is one of the GOODS-S candidates. 
Furthermore, the brightest of our candidates (at $z=10.2\pm0.4$) is robustly detected also at 3.6\,$\mu$m ($6.9\sigma$), revealing a flat UV spectral energy distribution with a slope $\beta=-2.0\pm0.2$, consistent with demonstrated trends with luminosity at high redshift.
The abundance of such luminous candidates suggests that the luminosity function evolves more significantly in $\phi_*$ than in $L_*$ at $z\gtrsim8$ with a higher number density of bright sources than previously expected.
Despite the discovery of these luminous candidates,
the cosmic star formation rate density for galaxies with SFR $>0.7$ \msol~yr$^{-1}$ shows an order-of-magnitude increase in only 170 Myr from $z \sim 10$ to $z \sim 8$, consistent with previous results given the dominance of low-luminosity sources to the total SFR density.
Based on the IRAC detections, we derive galaxy stellar masses at $z\sim10$, finding that these luminous objects are typically $10^9$ \msol. 
This allows for a first estimate of the cosmic stellar mass 
density at $z\sim10$ resulting in $\log_{10}\rho_{*} = 4.7^{+0.5}_{-0.8}$ \msol~Mpc$^{-3}$ for
galaxies brighter than $M_{UV}\sim-18$.
The remarkable brightness, and hence luminosity, of these $z\sim9-10$ candidates highlights the opportunity for deep spectroscopy to determine their redshift and nature, demonstrates the value of additional search fields covering a wider area to understand star-formation in the very early universe, and highlights the
opportunities for JWST to map the buildup of galaxies at redshifts
much earlier than $z\sim10$.
\end{abstract}


\section{Introduction}

The very sensitive near-infrared imaging with the Wide Field Camera 3 (WFC3/IR) on board
the Hubble Space Telescope ($HST$) has enabled the extension of the
observational frontier for galaxies to beyond $z\sim9$, only 500 Myr after
the Big Bang.  However, detecting galaxies at such redshifts is clearly
approaching the limit of what is possible with the $HST$.  Despite
extremely deep imaging over the Hubble Ultra-Deep Field
\citep[HUDF;][]{Beckwith06,Illingworth13,Ellis13}, only one reliable
$z\sim10$ galaxy candidate could be identified over this small field and
over additional wider area blank field data in the Chandra Deep Field South region
\citep{Oesch13}. 
The first reported $z\sim10$ galaxy, UDFj-39546284 \citep{Bouwens11a}, now has an uncertain redshift based on newer data \citep{Ellis13,Brammer13,Bouwens13b}. 
Two more $z\sim10$ sources were detected in the Cluster
Lensing and Supernova survey with Hubble \citep[CLASH;][]{Postman12},
making use of lensing magnification of massive foreground clusters
\citep{Zheng12,Coe13}. 

As exciting as these detections are, the small number of $z>9$ galaxy
candidates makes it quite difficult to reliably determine the cosmic
star-formation rate density (SFRD) at these early times.  In particular,
does the cosmic SFRD increase slowly with time at $z>8$, as seen at  $z<8$,
or does it change more rapidly and dramatically as some models suggest?
While some authors \citep[e.g.,][]{Coe13,Ellis13} obtained results
consistent with the SFRD continuing the same steady decline from $z\sim8$
to $z\sim10$ as observed at lower redshifts,
the most extensive $z\sim10$ galaxy search to date \citep{Oesch13} found a
significant drop in the SFRD by about an order of magnitude from $z\sim8$
to $z\sim10$ when combining all published measurements at $z>8$.  Clearly,
enlarging the sample of $z>8$ galaxies would help to establish the rate at
which the cosmic SFRD increased.  

With the completion of the Cosmic Assembly Near-Infrared Deep Extragalactic
Legacy Survey \citep[CANDELS;][]{Grogin11,Koekemoer11} in 2013 August, it is
now possible to extend the search area for $z\gtrsim8$ galaxies. In
particular, we will focus on the GOODS-N dataset in this paper, where
the CANDELS survey acquired F105W ($Y_{105}$) imaging data and where much
more extensive multi-wavelength optical data are available from the Great Observatories Origins Deep Survey
\citep[GOODS][]{Giavalisco04a} than for the other CANDELS-Wide fields. As we
have demonstrated in previous papers, the limits placed on flux
measurements at shorter wavelengths play a crucial role in enhancing the
reliability of high-redshift galaxy searches by removing probable
low-redshift contaminants. The non-detections at wavelengths below the
Lyman break (i.e., for $z\sim9-10$ candidates the optical data together with the $Y_{105}$ imaging)
greatly lessen the problem of contamination by lower redshift sources.

With the detection of galaxies at $z>8$, a new 
challenge has become to characterize the physical properties of galaxies at 450-650 Myr.
A key parameter for any characterization is the stellar mass. This is
possible with the use of the $Spitzer$ Infrared Array Camera (IRAC), which samples rest-frame optical
light even for $z\sim10$ galaxies with its 4.5 \micron\ channel.
Unfortunately, the extremely faint HST $z\sim9-10$ galaxy candidates that were
identified down to $H_{160,AB}\sim30$ mag in the HUDF field are out of
reach of IRAC.  The two lensed candidates of \citet{Coe13} and
\citet{Zheng12} are sufficiently bright that they show weak
($\sim2$-$3\sigma$) IRAC detections, but the uncertainties are so large
that the stellar mass estimates are still not very reliable.  In the present paper, we
search for bright $z\sim9-10$ galaxy candidates in the GOODS-N field
where very deep $Spitzer$/IRAC data are  available. Such deep
$Spitzer$/IRAC data would allow for a first estimate of the galaxy stellar
mass density at $z\sim10$.

This paper is an extension of our previous analyses of the GOODS-S data
set and our $z\sim10$ Lyman Break galaxy (LBG) search
\citep{Bouwens11a,Oesch12a,Oesch13}. It is organized as follows: we
describe the data used for this analysis in Section \ref{sec:data} and
present our high-redshift candidate selection in Section
\ref{sec:selection}. These candidates are subsequently used in Section
\ref{sec:results} to derive new constraints on the evolution of the UV
bright galaxy population out to $z\sim10$.  Section \ref{sec:IRACDet}
provides an analysis of the stellar mass density at $z\sim10$ based on
robust IRAC detections.  Section \ref{sec:summary} summarizes our results
and briefly discusses the implications of our findings for planning future
surveys such as with the James Webb Space Telescope (JWST). In an appendix we note 
the outcome of a search for bright $z\sim9-10$ candidates in GOODS-S that was 
motivated by the discovery of such sources in GOODS-N.

Throughout this paper, we will refer to the HST filters F435W, F606W, F775W, F814W, F850LP, F105W, F125W, F140W, F160W as \bFilter, \vFilter, \iFilter, $I_{814}$, \zFilter, \yFilter, \jFilter, \jhFilter, \hFilter, respectively. Magnitudes are given in the AB system \citep{Oke83}, and we adopt a standard cosmology with $\Omega_M=0.3, \Omega_\Lambda=0.7, H_0=70$ kms$^{-1}$Mpc$^{-1}$, i.e. $h=0.7$, consistent with the most recent measurements from Planck \citep{Planck13XVI}.

\begin{deluxetable*}{lccccccccccccc}
\tablecaption{The 5$\sigma$ Depths of the Observational Data Used in this Paper\label{tab:data}}
\tablewidth{0 pt}
\tablecolumns{0}
\tablehead{\colhead{Field} & Area [arcmin$^2$] & \bFilter &\vFilter &\colhead{\iFilter} & \iwFilter &  \colhead{\zFilter}  &\colhead{$Y_{105}$} & \colhead{\jFilter} & \jhFilter & \colhead{\hFilter} & \colhead{K-band\tablenotemark{a}} & \colhead{IRAC 3.6} & \colhead{IRAC 4.5}  }
\startdata


 GOODSN-Deep & 64.5 & 28.0  & 28.2  & 27.6 & 28.9  & 27.7  & 27.6  & 28.1 & 26.8 & 27.8   & 25.0-26.6  & 27.0 & 26.7\\
GOODSN-Wide & 69.4 & 28.0  & 28.2  & 27.6 &  28.2  & 27.7  & 27.2  & 27.2 & 26.8 & 27.1    & 25.0-26.6  & 27.0 & 26.7

\enddata

\tablecomments{Depths are measured in circular apertures and are corrected to total fluxes using the flux growth curves of stars. The aperture diameters were
0\farcs35 for the ACS and WFC3 data, 0\farcs6 for the K-band,  and 2\farcs0 for the IRAC data. These were chosen to be consistent with the actual aperture sizes used for photometry of our sources of interest. }
\tablenotetext{a}{The depth of the MOIRCS K-band data varies significantly across the field due to non-uniform exposure times. }

\end{deluxetable*}

\section{Data}
\label{sec:data}

The dataset analysed in this paper consists of deep, high-resolution HST imaging covering 0.4-1.6 $\mu$m, in addition to ground-based K-band data, as well as Spitzer IRAC imaging at 3.6 and 4.5 $\mu$m. These datasets are discussed in the next section and a summary of their depths is listed in Table \ref{tab:data}.

\subsection{HST Data in GOODS-North}

We base this paper on the entire  WFC3/IR and $I_{814}$ Advanced Camera for Surveys (ACS) data over the GOODS-N
field from the completed CANDELS survey. The last data were taken on 2013
August 10. For details on the survey layout we refer to the CANDELS team
papers \citep{Grogin11,Koekemoer11}. Briefly, the CANDELS GOODS-N
field is part of the CANDELS-Deep survey, for which additional $Y_{105}$
imaging was obtained. The $Y_{105}$ imaging is not available in the general
CANDELS-Wide component. 

The central $\sim65$ arcmin$^2$ of the GOODS-N field was covered by $\sim5$
orbits of WFC3/IR imaging data in $J_{125}$ and $H_{160}$ reaching to 27.8
mag ($5\sigma$) and by $\sim3$ orbits of $Y_{105}$ imaging reaching  27.6
mag. Furthermore, two flanking fields totaling  $\sim70$ arcmin$^2$ of the
CANDELS-Wide program completed the WFC3/IR coverage of the GOODS-N
field with roughly $1$ orbit in each of the three filters, $Y_{105}$,
$J_{125}$, and $H_{160}$, which results in a depth of $27.0-27.2$ mag in
all three filters.  We also include all the \jhFilter\ imaging data
available over GOODS-N. These are very shallow exposures, mostly being used
as pre-imaging for the GOODS-N grism program (GO:11600, PI: Wiener) with an
exposure time of 800s, in addition to a few supernova follow-up
observations from the CANDELS survey. Nevertheless, these can be useful for
spectral energy distribution (SED) analyses of brighter sources.

We downloaded all the individual WFC3/IR data from the Barbara A. Mikulski Archive for Space Telescopes (MAST) and the Canadian Astronomy Data Centre (CADC) and reduced the data using standard procedures, as described in detail in \citet{Illingworth13}.
We used the persistence masks provided by the Space Telescope Science Institute (STScI) to mask all pixels significantly affected by persistence from previous exposures.
We registered all WFC3/IR frames to the official v2 GOODS-N ACS $z_{850}$-band data before drizzling to a mosaic with final pixel size of $0\farcs06$ and a tangent-plane projection aligned to the GOODS-N ACS data.
The RMS maps produced by multidrizzle were rescaled to match the actual fluctuations present in the data as measured through circular apertures of 0\farcs35 diameter randomly placed on empty sky positions in the images. 

The ACS data used here are a somewhat deeper reduction than the publicly
released v2 GOODS-N imaging.  We included the additional data from supernova
follow-up programs \citep[see][]{Bouwens07}. Furthermore, we reduced the
new $I_{814}$-filter data obtained as part of parallel imaging from the
CANDELS program. The reduction includes corrections for charge transfer
inefficiency and was performed analogously to the eXtreme Deep Field (XDF) data reduction \citep{Illingworth13}. The depth of these $I_{814}$ images
surpasses all other GOODS ACS images, reaching a $5\sigma$ limit of
$I_{814} = 28.2-28.9$ mag.

The angular width of the point-spread function  of the data used here is $\sim0\farcs09$ and $\sim0\farcs16$ for the ACS and WFC3/IR imaging, respectively, as measured from unsaturated stars in the field.  


\subsection{The IRAC Data Set}

From previous $z>8$ analyses it became  clear that longer-wavelength constraints from $Spitzer$/IRAC are essential in order to remove contamination from dusty, intermediate redshift sources \citep[e.g.,][]{Oesch12a}. Furthermore, IRAC samples the rest-frame visible at $z>4$, which is crucial for stellar mass constraints. We therefore include all the IRAC data that are available over the GOODS-N region as part of several programs. In particular, we analyzed the 3.6\,$\mu$m and 4.5\,$\mu$m channel IRAC reductions from the Spitzer Extended Deep Survey (SEDS) and S-CANDELS team \citep[PI: Fazio; see also][]{Ashby13}. 

The Spitzer S-CANDELS program (P.I.\ G.~Fazio) is a Cycle 8 \Sp\ Exploration
Science project to map $\sim$0.2~deg$^2$ in the five CANDELS fields
to 50~hours depth with IRAC, thereby reaching magnitude $\sim$26.8 at 3.6
and 4.5~\micron.  Data in GOODS-N were obtained in two epochs in 2012
and combined with pre-existing IRAC data from SEDS \citep{Ashby13}
and the original GOODS program \citep{Dickinson03}.  The corrected, basic calibrated data (cBCD) frames
from the \Sp\ archive were combined into calibrated mosaics following
the same procedures as for SEDS \citep{Ashby13}. The achieved depth
at the positions of the $z\sim9$ candidates (Sec.~3) is $\sim$50~hr
in both IRAC bands except that the 4.5~\micron\ observation of GN-z10-1 has $\sim$72~hr. In blank sky areas the average IRAC 5$\sigma$ depths are 27.0 and 26.7 mag (within 1\arcsec\ radius apertures) in the two IRAC channels. Further analysis of these mosaics, including catalog
creation and completeness estimates, is ongoing and will be reported
elsewhere (Ashby et al., in preparation).

A significant challenge of the IRAC data is the poor angular resolution,
${\rm FWHM} = 1\farcs7$ \citep{Fazio04}, leading to source confusion.
This can be largely overcome, however, through neighbor subtraction based
on the high-resolution HST data. This is particularly important for the
faint sources that we study here. 
We used an approach outlined in several previous papers \citep[see
e.g.,][]{Labbe06,Grazian06,Laidler07} that models a region around a source
of interest using its $HST$ $H_{160}$ image convolved to IRAC resolution
and subtracts all neighbors to give a "cleaned" source. Subsequently we
measured fluxes in $1\arcsec$ radius circular apertures and multiplied by a
factor 2.4-2.6 to correct for light outside the aperture. The variation in
the aperture correction is due to variations in the position dependent IRAC
PSF as measured from nearby stars in the field.

While subtraction of the flux from the neighboring sources does not always
work if a candidate is too close to a very bright foreground source, it is
very effective in the majority of cases. The modeling and subtraction
approach has been refined extensively and now allows us to perform clean
IRAC photometry for  $\sim75$-$80\%$ of sources in the field, a factor $\sim2\times$
larger than without neighbor subtraction.


\subsection{MOIRCS K-band Data over GOODS-N}

In order to bridge the gap in wavelength range between the 1.6\,\micron\ 
probed by $HST$ and the 3.6\,\micron\ covered by IRAC, we also made use of a
deep K-band stack over GOODS-N from the multi-object infrared camera and spectrograph for Subaru
\citep[MOIRCS][]{Kajisawa06,Bouwens08}. These data were reduced using the standard
MOIRCS pipeline procedures. The average seeing measured from the final
stack of the data is 0\farcs55. Over the region of interest, this stack
varies in depth between 25.0 and 26.6 mag AB, as measured in small
circular apertures of 0\farcs6 diameter. Total magnitudes were derived
using aperture corrections computed from the profiles of a few bright,
non-saturated stars in the field.

\subsection{Supporting HST Fields: HUDF09/12/XDF and GOODS-South}

In the last part of this paper, we will derive new constraints on the UV
luminosity function (LF) and the cosmic SFRD at $z\sim10$ based on the
largest possible dataset. We will therefore combine the GOODS-N CANDELS
dataset with an identical analysis over previous deep $HST$ imaging fields.
In particular, we directly include the $z\sim10$ search and analysis from
the HUDF09/12/XDF and GOODS-S field from our previous paper
\citep{Oesch13}. However, motivated by the bright galaxies in GOODS-N, we systematically re-analyzed the GOODS-S dataset with detection criteria that are better matched to those used in GOODS-N. The result of this is presented in the appendix.
 The combined WFC3/IR+ACS dataset spans an area of
$\sim300$ arcmin$^2$ and ranges in depth from $H_{160}=27.5-30.0$ mag AB
($5\sigma$). 

\begin{figure*}[tbp]
	\centering
	\includegraphics[width=0.42\linewidth]{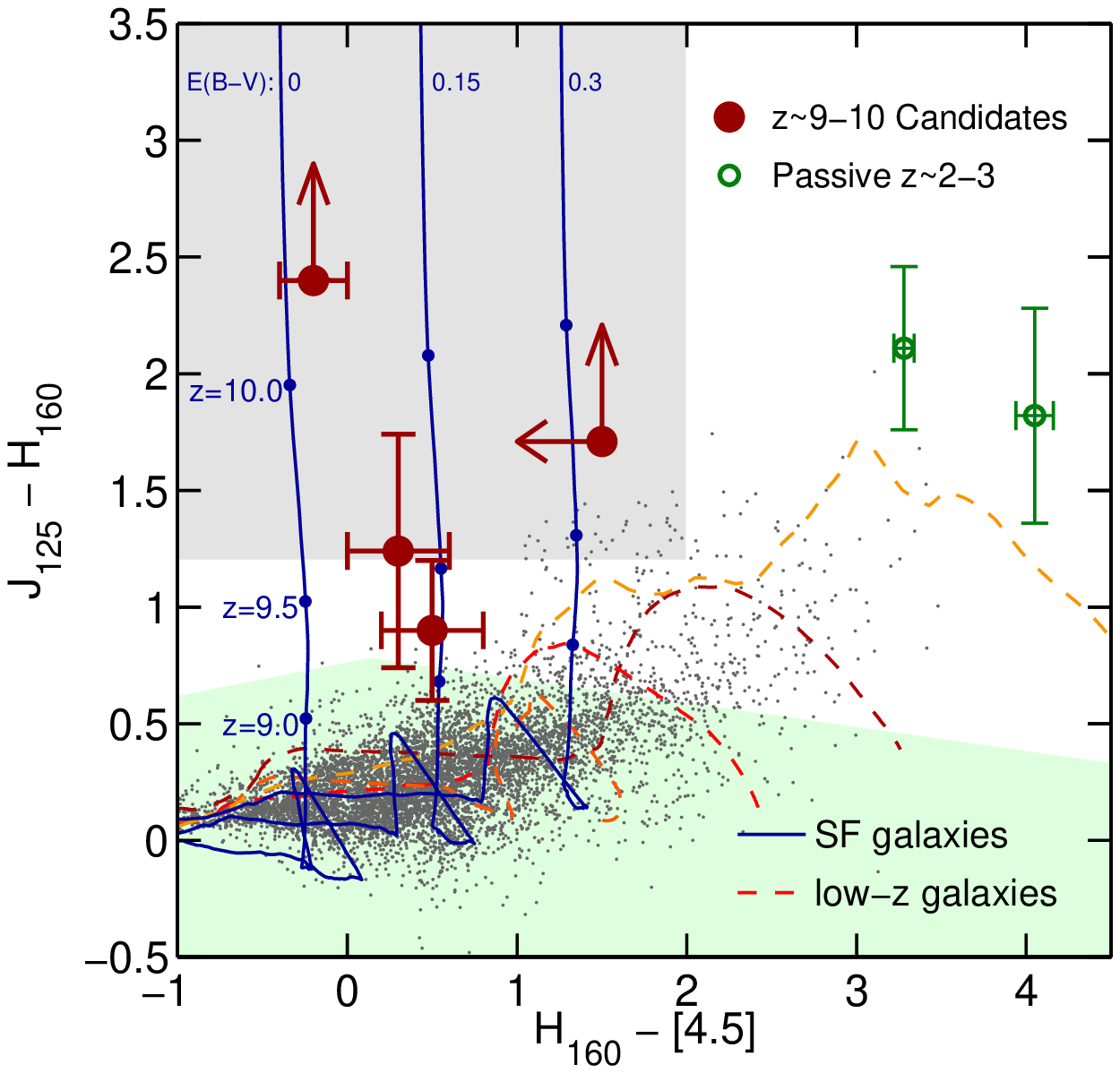} 
\includegraphics[width=0.42\linewidth]{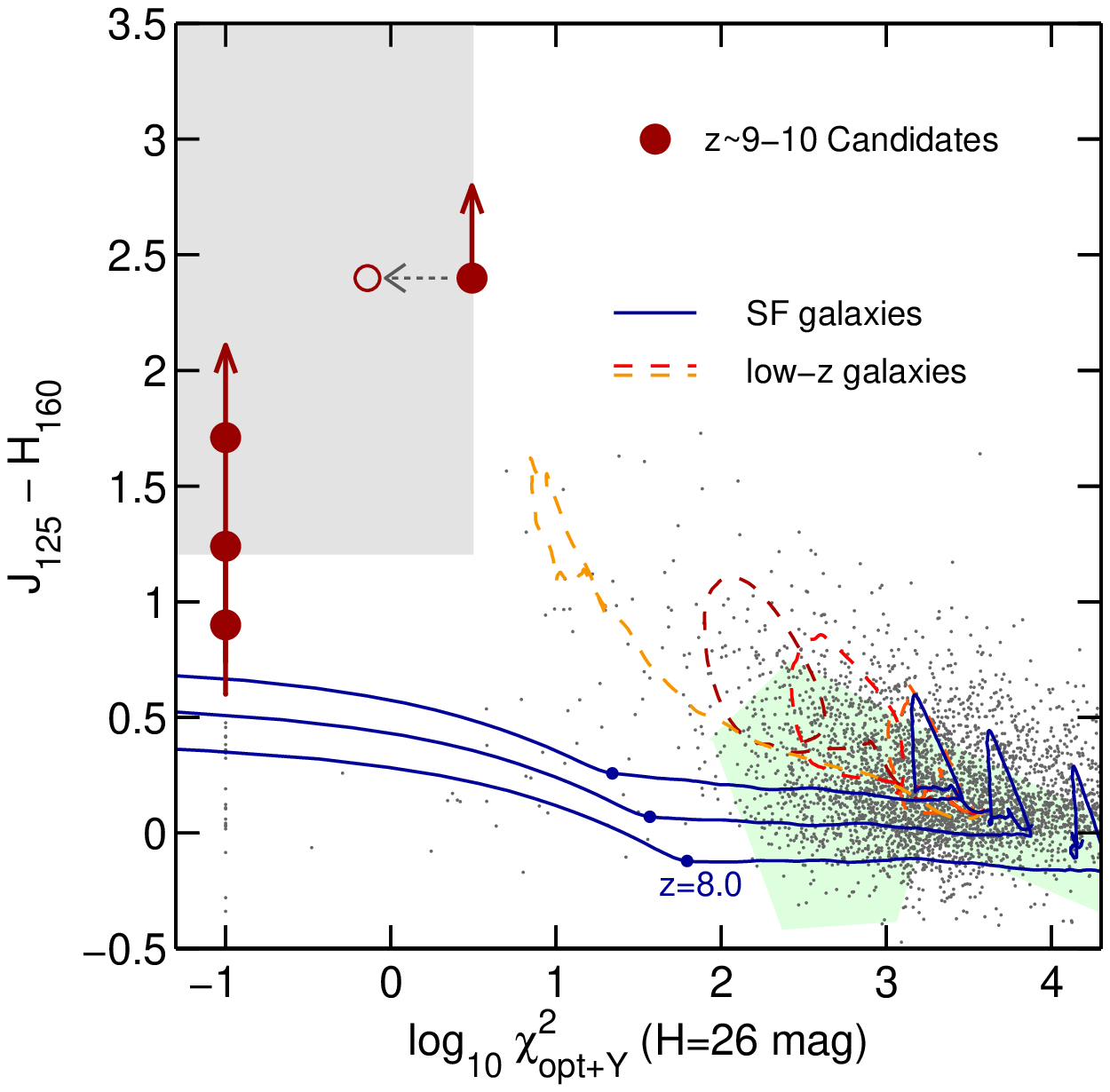} 
  \caption{\textit{Left -- } The $J_{125}-H_{160}$ vs $H_{160}-[4.5]$ color diagram showing the location of the GOODS-N $z>9$ galaxy candidates relative to other, lower redshift galaxy SED tracks. The four $z\sim9-10$ candidates are shown with dark red circles with $1\sigma$ errorbars on their colors. For non-detections, 2$\sigma$ color limits are shown. The only other sources identified as having no optical detections are shown as green circles. These two sources are very bright in IRAC, consistent with being dusty or passive sources at $z\sim2-3$. They are thus not included in our subsequent analysis. Also shown as small points are the locations of all sources with reliable IRAC 4.5\,\micron\ and $H_{160}$ flux measurements ($>10\sigma$) from the CANDELS GOODS-S catalog of \citet{Guo13}. These sources nicely follow the redshift tracks of evolved $z<5$ SEDs (dashed yellow to red lines). The gray shaded area indicates the region in color-color space expected for $z\geq9.5$ star-forming galaxies with $J_{125}-H_{160}>1.2$. Note that the initial selection of the four GOODS-N candidates was $J_{125}-H_{160}>0.5$, which selects sources at $z\gtrsim9$. The blue lines indicate the tracks of star-forming galaxies with different amounts of dust extinction ($\mathrm{E(B-V)}=0,~0.15,~0.3$).  Visual inspection and SED fits of the few sources from the CANDELS catalog that lie in the gray area show that they are all relatively compact, intermediate redshift, passive sources. Unlike the real high-redshift candidate sources, these are significantly detected in the optical data and so they can be clearly rejected as contaminants (see the figure to the right). The green shaded area that is well-separated from the gray selection zone indicates the area where Galactic stars are expected, including very low mass M, L, T, and Y dwarfs \citep[see also][]{Coe13}. 
  \textit{Right -- } Plot of $\chi^2_{opt+Y}$ (see Section 3.1) against the $J_{125}-H_{160}$ color, representing our second selection criterion for $z>9$ galaxies (which minimizes low redshift contaminants). The four $z\sim9-10$ candidates are again shown with dark red circles, while small gray dots indicate galaxies with 10$\sigma$ $H_{160}$ detections in our GOODS-N CANDELS-Wide catalog. Sources with $\log_{10}\chi^2_{opt+Y}<-1$ are limited at that value. The shaded areas and lines represent the same as in the left panel. The $\chi^2_{opt+Y}$ values for the SED tracks were normalized to $H_{160}=26$ mag and were computed for the CANDELS-Wide field depth. The gray arrow and the open circle shows how the  value of $\chi^2_{opt+Y}$ changes for GN-z10-1 if the $z_{850}$-band is excluded in which a background fluctuation causes a positive 1.5$\sigma$ flux measurement.
  }
	\label{fig:colcol}
\end{figure*}

\begin{figure*}[htp]
	\centering
	\includegraphics[width=0.8\linewidth]{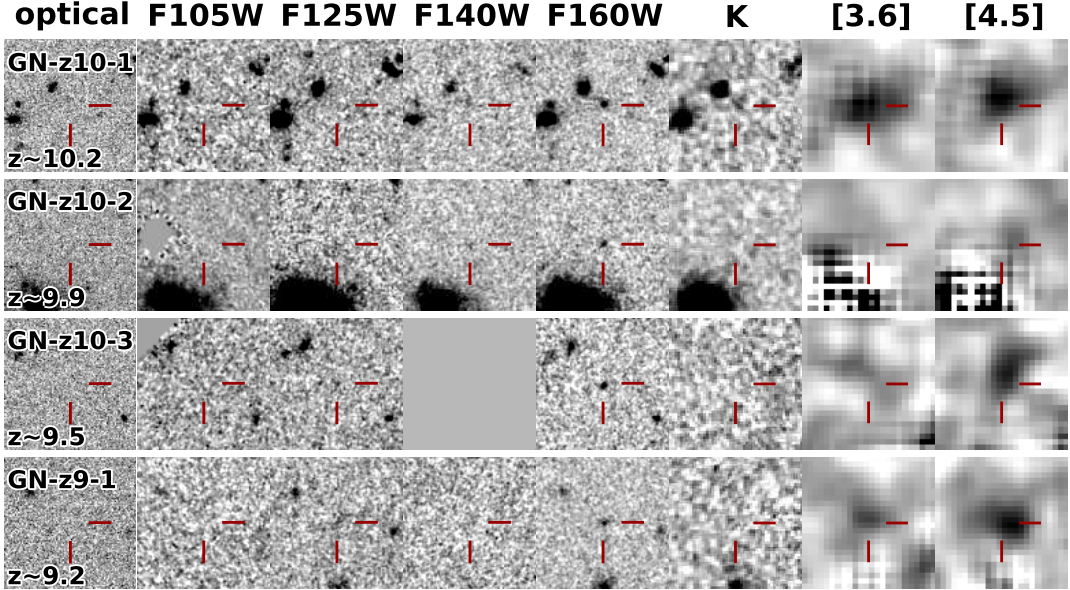}
  \caption{6\arcsec$\times$6\arcsec\ images of the four $z\geq9$ galaxy candidates identified in the CANDELS GOODS-N data. From left to right, the images show a stack of all optical bands, \yFilter, \jFilter, $JH_{140}$, \hFilter, MOIRCS K, and neighbor-subtracted IRAC 3.6\,$\mu$m and 4.5\,$\mu$m images. The stamps are sorted from high to lower photometric redshift from SED fits (indicated in the lower left, see also Table \ref{tab:candidates}). The IRAC neighbor-subtraction works well for all sources except for GN-z10-2, where the nearby foreground source is too bright, and clear residuals are visible at the location of the candidate. Only IRAC upper limits are therefore included for this source in the following analysis. 
Clearly, all other sources show significant ($>4.5\sigma$) detections in the 4.5\,$\mu$m channel. The brightest source (GN-z10-1) is also detected at 6.9$\sigma$ in the 3.6\,$\mu$m channel.
  With the exception of the brightest candidate, which is weakly detected in the K-band (at 2$\sigma$), the MOIRCS K-band data provide only upper limits.}
	\label{fig:stampszgtr8}
\end{figure*}

\section{The GOODS-North $z>9$ Galaxy Sample}
\label{sec:selection}

\subsection{Sample Selection}

The identification of Lyman Break galaxies in the epoch of reionization makes use of the almost complete absorption of UV photons shortward of the redshifted Ly$\alpha$ line due to a high neutral hydrogen fraction in the inter-galactic medium. At $z>9$ the Ly$\alpha$ absorption shifts into the $J_{125}$ band, which renders star-forming galaxies red in their $J_{125}-H_{160}$ colors. Our initial selection criterion is therefore to search for galaxies with $J_{125}-H_{160}>0.5$ and non-detections in the shorter wavelength data. In the second part of this paper, we will restrict the sample to a more conservative criterion with $J_{125}-H_{160}>1.2$, which includes only galaxies at $z\gtrsim9.5$. This selection $J_{125}-H_{160}>1.2$ also matches our previous GOODS-S analysis and allows us to use a larger sample for the subsequent analysis.

Source catalogs were obtained with SExtractor \citep{Bertin96}, which was run
in dual image mode with the $H_{160}$-band as the detection image. 
All images were convolved to the $H_{160}$ point-spread function when performing photometry, and colors were measured in small Kron apertures (Kron factor 1.2),
typically 0\farcs2 radius. 
Total magnitudes were derived from larger elliptical apertures using the standard Kron
factor of 2.5, typically 0\farcs4 radius, with an additional correction to total fluxes based on
the encircled flux measurements of stars in the $H_{160}$ band. This last correction was typically $\sim0.2$ mag but depended on the actual Kron aperture size of individual galaxies.

Based on these catalogs, the following $HST$ selection criteria were applied:
\begin{equation}
	(J_{125}-H_{160})>0.5
\end{equation}
\[
S/N(B_{435} \mathrm{ ~to~ } Y_{105})<2\quad \wedge \quad \chi^2_{opt+Y}<3.2
\]
in addition to at least 5$\sigma$ detections in $H_{160}$ (see Figure \ref{fig:colcol}).  The
$\chi_{opt+Y} ^2$ for each candidate source was computed as $\chi_{opt+Y}
^2 = \Sigma_{i} \textrm{SGN}(f_{i}) (f_{i}/\sigma_{i})^2$
\citep{Bouwens11c} where $f_{i}$ is the flux in band $i$ in a
consistent aperture, $\sigma_i$ is the uncertainty in this flux, and
SGN($f_{i}$) is equal to 1 if $f_{i}>0$ and $-1$ if $f_{i}<0$, and the
summation is over the \bFilter, \vFilter, \iFilter, $I_{814}$,
\zFilter, and \yFilter\ bands.
The limit of $\chi^2_{opt+Y}=3.2$ was chosen to result only in a small reduction in the selection volume of real $z>9$ sources (20\% based on Gaussian statistics), while 
 efficiently excluding lower redshift contamination (see the right panel of Figure \ref{fig:colcol}). This reduction in the selection volume is accounted for in our calculations of the UV LF in Section \ref{sec:results}.

These HST selection criteria resulted in a total of six potential candidates in the full GOODS-N WFC3/IR dataset. However, two of these sources are extremely bright in the $Spitzer$/IRAC bands with $H_{160}-[4.5]>3.2$. 
As shown in the left panel of Figure \ref{fig:colcol}, this is much redder than expected for a real high-redshift galaxy. However, it is consistent with similarly red sources with photometric redshifts $z\sim2-4$ that we identified as contaminants in our GOODS-S high-redshift search \citep{Oesch12a,Oesch13}. Such sources are very interesting for $z\sim2-4$ studies \citep[see, e.g.,][]{Huang11,Caputi12}, but will not be discussed further here. We exclude these two sources from our analysis and proceed with only four potential $z>9$ galaxy candidates.

\begin{figure}[tbp]
	\centering
\includegraphics[width=1.02\linewidth]{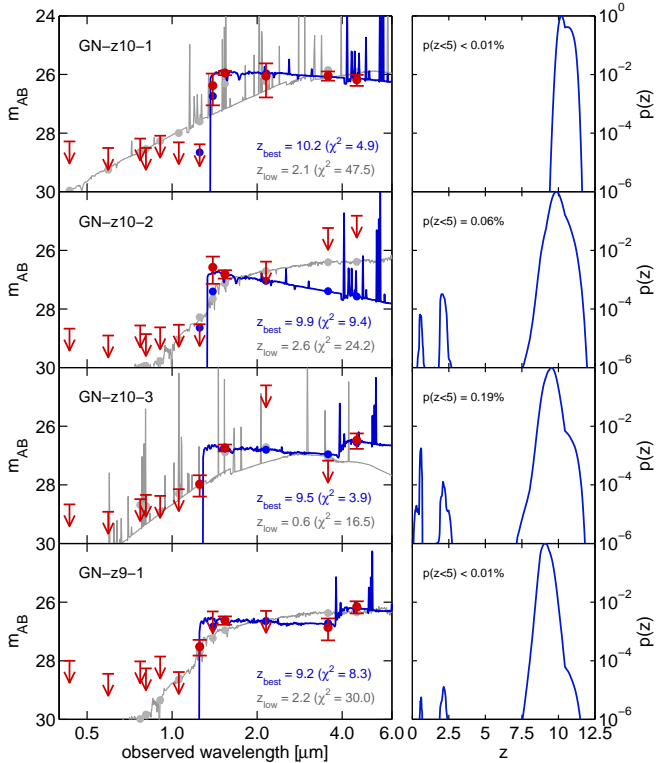}	  
  \caption{Spectral energy distribution fits to the HST and Spitzer/IRAC photometry of the four GOODS-N $z\sim9-10$ galaxy candidates (\textit{left}) together with the redshift likelihood functions (\textit{right}). The measurements and their upper limits ($2\sigma$) are shown in dark red. 
Best-fit SEDs are shown as blue solid lines, in addition to the best low redshift solutions in gray. The corresponding SED magnitudes are shown as filled circles. For all sources, the $z\geq9$ solution fits the observed fluxes significantly better than any of the possible low-redshift SEDs.  
The integrated likelihoods for $z_\mathrm{phot}<5$ are all $<0.2\%$ as shown by the labels in the right panels. }	
\label{fig:sedFits}
\end{figure}

While three out of the four remaining sources show negative values of $\chi^2_{opt+Y}$, the brightest candidate (GN-z10-1) lies very close to the selection limit (see right panel of Figure \ref{fig:colcol}).
This is mostly driven by a 1.5$\sigma$ positive flux measurement in the \zFilter -band. From a visual inspection of that image, however, it appears that this is due to a feature in the background and is not associated with real flux from the source. 
If we remove this band, the $\chi^2_{opt+Y}$ drops to 0.7.
Nevertheless, the $\chi^2_{opt+Y}$ near the selection limit indicates that this source could be a potential contaminant. Based on spectral energy distribution (SED) fitting, however, we will show later in Section \ref{fig:sedFits} that no low redshift galaxy SED or stellar SED that we know of can reproduce the very red $J_{125}-H_{160}$ color break of this source together with its flat continuum longward of 1.6\,\micron. Taken together, these results suggest that the most likely
interpretation is that GN-z10-1 is at high redshift.

Stamps of the four viable high-redshift candidates are presented in Figure \ref{fig:stampszgtr8}, and their positions and photometry are listed in Tables \ref{tab:candidates} and \ref{tab:flux}.
As is evident from Fig. \ref{fig:stampszgtr8}, the four sources are all detected at $\geq7\sigma$ in the $H_{160}$ band. The brightest source is 15$\sigma$. Furthermore, all sources are seen in observations at other wavelengths, albeit at lower significance. With the exception of GN-z10-1, all show weak detections in $J_{125}$, and two are even seen weakly in the very shallow \jhFilter\ data.
Furthermore, the brightest source is detected at $2\sigma$ in the ground-based K-band data.

Neighbor-subtraction was applied to the IRAC data of all four $z\gtrsim9$ galaxy candidates. The resulting cleaned IRAC images are shown in the two right-hand columns of Figure \ref{fig:stampszgtr8}. As can be seen, three of these sources are  detected in at least one IRAC band. 
For source GN-z10-2, the residuals of the bright foreground neighbor are still visible, and its IRAC flux measurements are therefore highly uncertain. In order to provide some photometric constraints for this source from IRAC, we use conservative upper limits based on the RMS fluctuations in the residual image at the position of the bright foreground source. All flux measurements for these sources, together with the uncertainties are listed in Table \ref{tab:flux}.

\begin{deluxetable*}{llcccccc}
\tablecaption{Coordinates and Basic Photometry of $z>9$ LBG Candidates in the GOODS-N Field\label{tab:candidates}}
\tablewidth{\linewidth}
\tablecolumns{8}

\tablehead{\colhead{Name} & \colhead{ID} & RA & DEC &\colhead{$H_{160}$}  &\colhead{$J_{125}-H_{160}$}  & \colhead{$H_{160}-[4.5]$}   &  \colhead{$z_{phot}$\tablenotemark{$^\ddagger$}} }

\startdata

GN-z10-1 & GNDJ-625464314 & 12:36:25.46  &  +62:14:31.4  & $25.95 \pm 0.07$  & $>2.4$  & $-0.2 \pm 0.2$& 10.2$\pm0.4$\\  %
GN-z10-2 & GNDJ-722744224 & 12:37:22.74  &  +62:14:22.4  &$26.81 \pm 0.14$  & $>1.7$  & $(<1.5)^{\dagger}$  & 9.9$\pm0.3$ \\  
GN-z10-3 & GNWJ-604094296 &  12:36:04.09  &  +62:14:29.6  &$26.76 \pm 0.15$  & $1.2 \pm 0.5$  & $0.3 \pm 0.3$ & 9.5$\pm$0.4 \\ 
\noalign{\vskip .7ex} \hline \noalign{\vskip 1ex}
GN-z9-1 & GNDJ-652258424* &  12:36:52.25  &  +62:18:42.4  &$26.62 \pm 0.14$  & $0.9 \pm 0.3$  & $0.5 \pm 0.3$  & 9.2$\pm0.3$

\enddata

\tablenotetext{$^\ddagger$}{Photometric redshifts listed here are derived with ZEBRA. The EAZY code and template set returns consistent redshifts within $\Delta z=0.1$.}

\tablenotetext{*}{The source GN-z9-1 does not satisfy the criterion $J_{125}-H_{160}>1.2$ and is therefore not included in our analysis of the UV LF and SFRD evolution at $z>8$ in Section \ref{sec:results}, for which we combine data from several previous analyses which used that stricter criterion. Nonetheless, this is considered to be a robust detection of a $z\sim9$ candidate galaxy.  It is excluded from the analysis only because of our intent to use consistent selection criteria for the overall sample analysis.}
\tablenotetext{$^{\dagger}$}{$3\sigma$ upper limit due to uncertainties in the neighbor flux subtraction.}

\tablecomments{Color limits are $2\sigma$. The numbers in the source IDs are a combination of the last 5 digits of the RA and the last 4 digits of the declination, which results in a unique name for all sources in the GOODS-N field.}

\end{deluxetable*}

\begin{deluxetable*}{lcccc}
\tablecaption{Flux Densities of $z>9$ LBG Candidates in the GOODS-N Field\label{tab:flux}}
\tablewidth{\linewidth}
\tablecolumns{5}

\tablehead{Filter  & GN-z10-1 &GN-z10-2 &GN-z10-3 &GN-z9-1 } 
\startdata 

%
%
%
  \bFilter & $7 \pm 9$  & $3 \pm 6$  & $-2 \pm 6$  & $-11 \pm 11$  \\ 
 \vFilter & $2 \pm 7$  & $-3 \pm 5$  & $-5 \pm 5$  & $2 \pm 8$  \\ 
 \iFilter & $5 \pm 10$  & $6 \pm 7$  & $6 \pm 7$  & $-9 \pm 11$  \\ 
 $I_{814}$ & $3 \pm 7$  & $1 \pm 5$  & $-2 \pm 8$  & $0 \pm 9$  \\ 
 \zFilter & $17 \pm 11$  & $-7 \pm 6$  & $-7 \pm 8$  & $-14 \pm 13$  \\ 
 \yFilter & $-7 \pm 9$  & $-7 \pm 7$  & $-2 \pm 10$  & $-18 \pm 8$  \\ 
 \jFilter & $11 \pm 8$  & $12 \pm 7$  & $23 \pm 8$  & $36 \pm 9$  \\ 
 \jhFilter & $102 \pm 47$  & $85 \pm 34$  & \nodata  & $86 \pm 54$  \\ 
 \hFilter & $152 \pm 10$  & $68 \pm 9$  & $73 \pm 8$  & $82 \pm 11$  \\ 
 K & $137 \pm 67$  & $-45 \pm 51$  & $85 \pm 261$  & $76 \pm 55$  \\ 
 IRAC 3.6\,$\mu$m & $139 \pm 20$  & $(<81)$\tablenotemark{*}   & $39 \pm  21$  & $65  \pm  18$  \\ 
 IRAC 4.5\,$\mu$m & $122 \pm 21$  & $(<119)$\tablenotemark{*}  & $93 \pm 21$  & $125  \pm 20$  

 \enddata

 \tablecomments{Measurements are given in nJy with $1\sigma$ uncertainties.}
 
\tablenotetext{*}{ $3\sigma$ upper limit due to uncertainties in the neighbor flux subtraction.} 
 
\end{deluxetable*}

\subsection{Photometric Redshift Analysis}
\label{sec:SEDanalysis}

Figure \ref{fig:sedFits} shows the SED fits to the fluxes of the four high-redshift galaxy candidates. These are derived with the photometric redshift code  ZEBRA \citep{Feldmann06,Oesch10c} using a large library of stellar population synthesis template models based on the library of \citet{Bruzual03}. Additionally, we added nebular line and continuum emission to these template SEDs in a self-consistent manner, i.e., by converting ionizing photons to  H and He recombination lines \citep[see also, e.g.,][]{Schaerer09}. Emission lines of other elements were added based on line ratios relative to H$\beta$ tabulated by \citet{Anders03}. 
The template library adopted for the SED analysis is based on both constant and exponentially declining star-formation histories of varying star-formation timescales ($\tau = 10^8$ to $10^{10}$ yr). All models assume a Chabrier initial mass function and a metallicity of 0.5$Z_\odot$, and the ages range from $t = 10$ Myr to 13 Gyr. However, only SEDs with ages less than the age of the Universe at a given redshift are allowed in the fit. Dust extinction is modeled following \citet{Calzetti00}.

As is evident in Figure \ref{fig:sedFits}, all candidates have a best-fit photometric redshift at $z\geq9$ with uncertainties of $\sigma_{z}=0.3$-$0.4$ (1$\sigma$). These uncertainties could be reduced with deeper \jhFilter\ imaging data in the future. 
However, the high redshift nature of these sources is quite secure given the current photometry. The right panels of Fig. \ref{fig:sedFits} show the redshift likelihood function. The integrated probability for $z<5$ solutions is strikingly small for each of these sources.  GN-z10-3 has the highest low-redshift likelihood with only 0.2\%. In this case, the best-fit low redshift SED is a combination of high dust extinction and extreme emission lines, which line up to boost the fluxes in the $H_{160}$ and IRAC 4.5\,$\mu$m bands. It is unclear how likely the occurrence of such an SED really is. However, deeper $Y_{105}$ data or spectroscopic observations could rule out such an SED. Some first spectroscopic constraints are already available from shallow WFC3 grism observations (see Section \ref{sec:grism}).

As a cross-check, we also tested and confirmed the high-redshift solutions with the photo-z code EAZY \citep{Brammer07}. In particular, we fit photometric redshifts with templates that include emission lines as included in the v1.1 distribution of the code\footnote{available at \url{http://code.google.com/p/eazy-photoz/}}. The best-fit EAZY redshifts are all within 0.1 of the ZEBRA values listed in Table \ref{tab:candidates}.

\subsection{Possible Sample Contamination}

As we will show in Section \ref{sec:NexpGOODSN}, the detection of such bright $z\sim9-10$ galaxy candidates in the GOODS-N dataset is  surprising given previous constraints on UV LFs at $z>8$. A detailed analysis of possible contamination is therefore particularly important. We discuss several possible sources of contamination in the next sections.

\subsubsection{Emission Line Galaxies}

Strong emission line galaxies have long been known to potentially
contaminate very high-redshift sample selection. These are a particular concern in datasets
which do not have very deep optical data to establish a strong spectral
break through non-detections \citep[see,
e.g.,][]{Atek11,vanderWel11,Hayes12}. Sources with extreme
rest-frame optical line emission may also contaminate $z\gtrsim9$ samples
if the $z\sim10$ candidate UDFj-39546284 \citep[][]{Bouwens11a,Oesch12a} is
any guide.  In that case, the extremely deep supporting data did not result
in any detection shortward of the $H_{160}$ band, but other evidence
(tentative detection of an emission line at 1.6\,$\mu$m and the high
luminosity of  UDFj-39546284) indicates that an extreme emission line galaxy at $z\sim2.2$ is
a more likely interpretation of the current data \citep[see][]{Bouwens13b,Ellis13,Brammer13,Capak13}.

In our SED analysis in Section \ref{sec:SEDanalysis}, we specifically
included line emission in order to test for  contamination from strong
emission line sources.  Indeed, for two of the candidates, the best-fit
low-redshift photometric redshift solutions are obtained from a combination
of extreme emission lines and high dust extinction. However, all candidates
are detected (although sometimes faintly) in several non-overlapping filters.
For example, with the exception of GN-z10-1, all sources show some
flux in the $J_{125}$ filter, as well as  a clear detection in $H_{160}$.
It is therefore unlikely that the detected $HST$ flux originates from
emission lines alone.  Furthermore, three of the four candidates show
robust detections in the IRAC bands, which further limits the likelihood of contamination by
pure line emitters. For example, GN-z10-1 (the brightest source),
shows evidence for a  flat continuum from the $HST$ $H_{160}$ to the IRAC
3.6\,$\mu$m and 4.5\,$\mu$m bands. As can be seen from Figure
\ref{fig:sedFits}, while this can be mimicked with the combination of
[\ion{O}{3}]/H$\beta$ contamination in the $H_{160}$ band and continuum
emission in the IRAC channels, the shorter wavelength flux limits rule out
such a lower redshift solution.  

Taken together, the likelihood that the sources here are lower-redshift
emission line galaxies is low.  The emission line constraints from the
IRAC filters are discussed further in Section \ref{sec:IRACDet} where we
present galaxy stellar mass estimates.

\begin{figure}[tbp]
	\centering
\includegraphics[width=\linewidth]{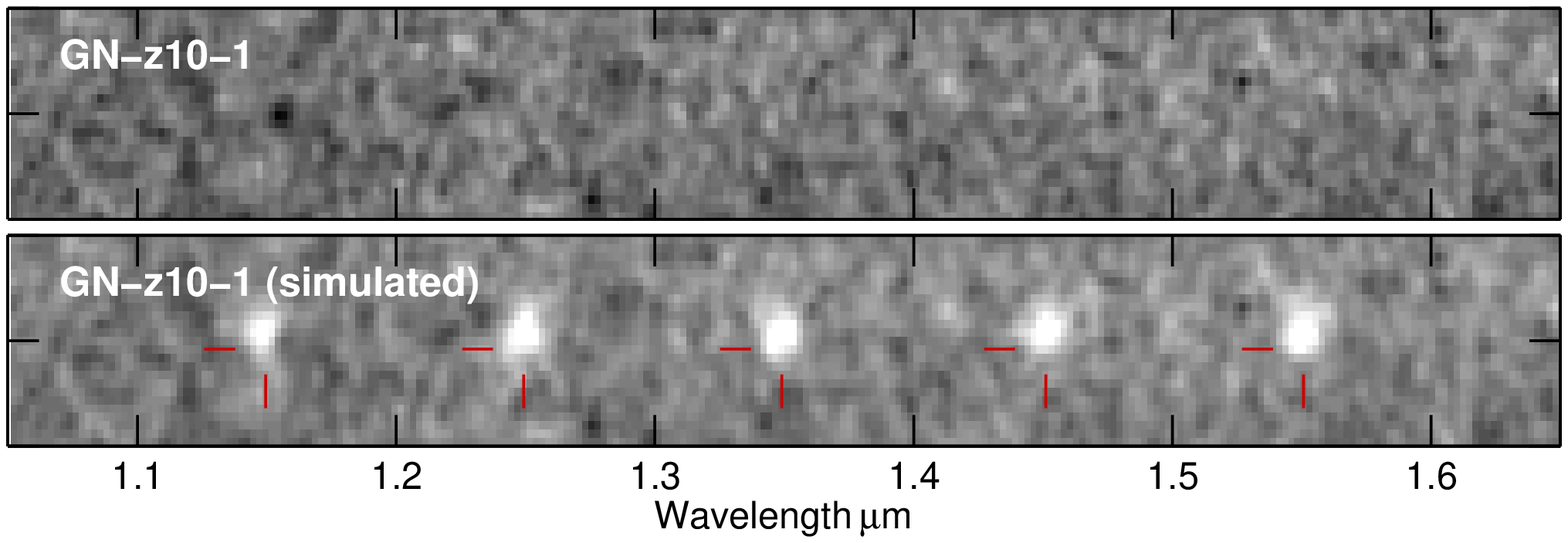}
\includegraphics[width=\linewidth]{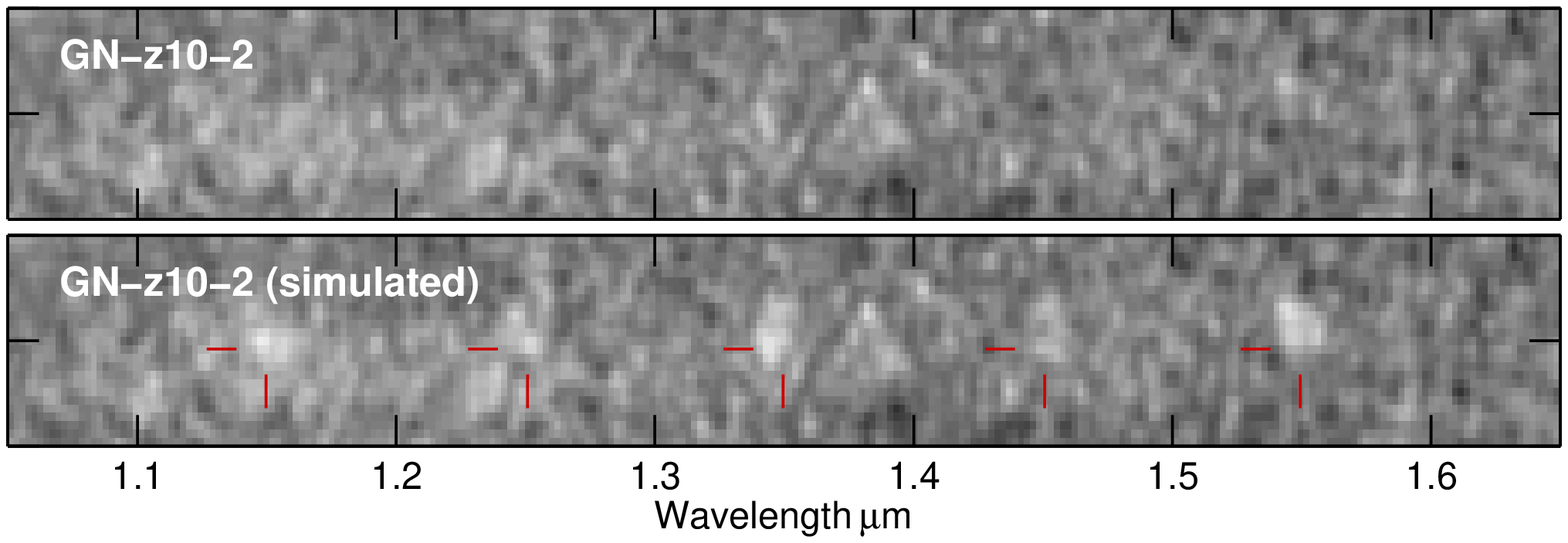}	 
\includegraphics[width=\linewidth]{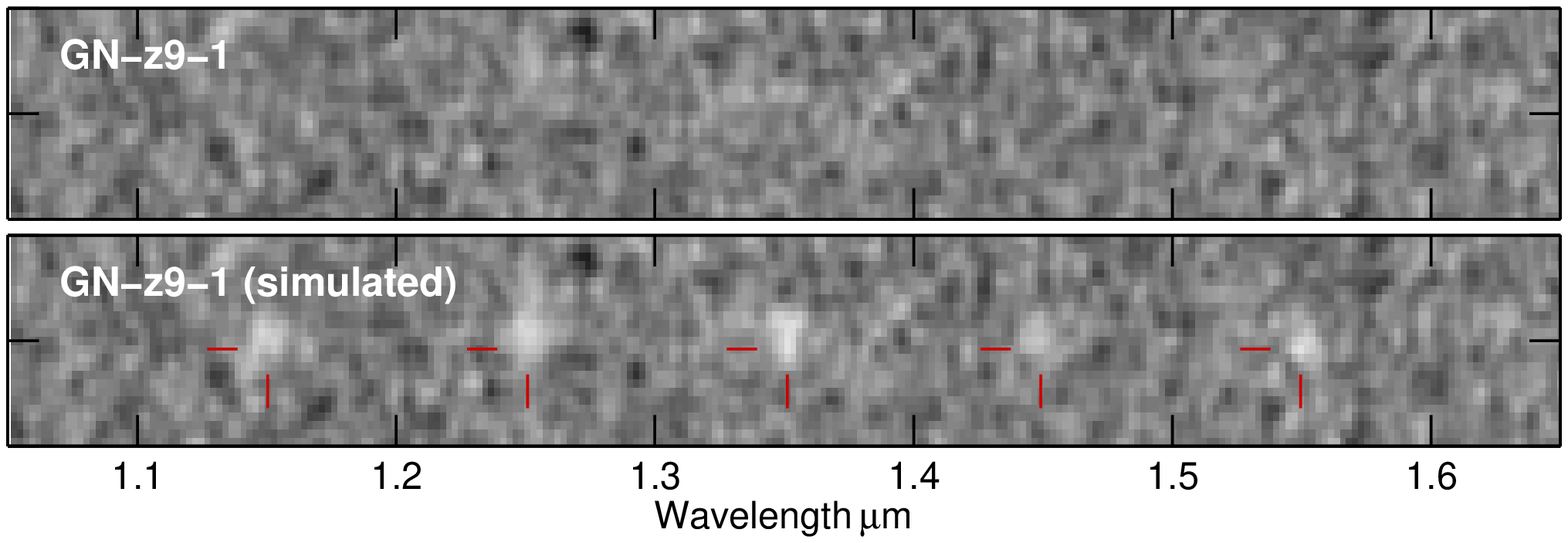}	 
  \caption{2D WFC3/IR grism G141 spectra for the three sources for which data are available. These are GN-z10-1 (top two panels), GN-z10-2 (middle two panels) and GN-z9-1 (bottom panels) as labelled in the plots. The spectra of these sources are expected to run along the center of each panel in the horizontal direction. The spectra were smoothed slightly with a Gaussian. No significant line emission is detected for any of the three sources.
   Below the original data, we show a panel with a simulation of pure emission line sources
at five different wavelengths, as indicated by red tick marks,   
    with a line flux corresponding to the $H_{160}$ photometry (5.5$\times10^{-17}$ \fluxunit\ for the brightest source, and 2.5$\times10^{-17}$ \fluxunit\ for the fainter two). Despite some residual contamination from a foreground source in the spectrum of GN-z10-2, such strong emission lines would have been significantly detected at $>4\sigma$. The grism data rule out pure emission line source contamination for these three sources.
   }
	\label{fig:grism}
\end{figure}

\subsubsection{Constraints from HST Grism Data}
\label{sec:grism}

Quantitative constraints on pure emission line sources can be obtained from the WFC3/G141 grism observations over GOODS-N from HST program 11600 (PI: Wiener). These spectra cover $\sim1.05-1.70~\mu$m at low resolution, reaching a $5\sigma$ emission line flux limit for compact sources of $\sim2-5\times10^{-17}$ erg~s$^{-1}$cm$^{-2}$ \citep[see][]{Brammer12}.
If the $H_{160}$-band flux originated from a single emission line, the observed magnitudes of our sources ($H_{160} = 26.0 - 26.8$ mag) would correspond to line fluxes of $2.5-5.5\times10^{-17}$ erg~s$^{-1}$cm$^{-2}$. These lines should thus be detectable as $\sim5 \sigma$ features.
We have therefore analyzed the grism spectra using reductions developed by the 3D-HST team. Three sources (excluding GN-z10-3) were covered with such data. 


The grism spectra were reduced using a newly developed pipeline by the 3D-HST team (Brammer et al.\ in prep.). 
The spectrum of GN-z10-1 was heavily affected by a varying sky background during the exposures, and it was necessary to manually exclude some of the affected readouts, which effectively reduced the total exposure time by less than 25\%.

The final 2D spectra of the three covered sources are shown in Figure \ref{fig:grism}. 
For sources GN-z10-1 and GN-z9-1, the spectra are blank, showing no significant features.
For source GN-z10-2, the spectrum is partially contaminated by the trace of a nearby bright source. Two extremely faint features may be visible at low significance ($\sim2\sigma$) at $\sim1.38$~\micron\ and 1.50~\micron. However, given the foreground contamination the data are currently inconclusive regarding these faint features. 

The spectra of these three sources provide very useful constraints on contamination by pure emission line sources.
Figure \ref{fig:grism} also shows simulated spectra of sources for which a single line could explain the whole flux in the $H_{160}$ filter, with lines at five different wavelengths. The simulated lines were computed based on the actually observed $H_{160}$-band profile of these sources in the spatial direction and assuming a Gaussian emission line with FWHM $=100$\,\AA\ in the dispersion direction. As can be seen, such a line would be significantly detectable even in the contaminated spectrum of GN-z10-2.

The grism data therefore rule out the contamination of a pure, single emission line source for these three candidates. However, based on the current data we can not rule out contamination by lower redshift sources with less extreme emission lines. Deeper data would be required to do so.
As we shall see, while any one constraint is not definitive, the grism emission line limits,
 the constraints on low-redshift contamination based on SED fitting,
and those from photometric scatter discussed in Sections \ref{sec:SEDanalysis} and \ref{sec:photscatter} indicate
that the sources identified here are highly likely to be at high redshift.

\subsubsection{Stellar Contamination?}

Stellar contamination can be a problem for high-redshift galaxy selections due to strong absorption features in dwarf stars, and the unusual brightness of our GOODS-N sources led us to give particular
attention to this aspect. We checked the surface brightness profiles of the sources, their colors, and their SEDs.  

For the brighter two sources GN-z10-1 and GN-z9-1, we measure
half-light radii of 0\farcs17 for both. After a simple correction for the
stellar PSF, these result in intrinsic half-light radii of only 0\farcs11.
At $z\sim10$, this would correspond to a physical size of only 0.5 kpc,
which, while small, is  consistent with the expectations from
extrapolating the $z\sim4-8$ size trends to $z\sim10$
\citep[e.g.,][]{Oesch10b,Ono12}.  Furthermore, both sources have
SExtractor full-width-at-half-maximum (FWHM) measurements more than
$1.8\times$ wider than for non-saturated stars in the $H_{160}$-band
imaging. The data suggest that these sources are resolved and so they are
unlikely to be stellar contaminants.  The fainter two sources
GN-z10-2 and GN-z10-3, while bright, are quite compact and are not detected with high enough S/N to rule out
stellar contamination purely based on their surface brightness profile. 



As Figure \ref{fig:colcol} shows, contamination by stars in our sample is
only expected from significant photometric scatter. Stars show colors of
$J_{125}-H_{160}\lesssim0.5$, even for very low mass dwarfs \citep[see
also][]{Oesch12a,Coe13}, which are typically the most important
contaminants in high-redshift samples. These colors are sufficiently blue
that our primary selection with $J_{125}-H_{160}>1.2$ is not expected to show
significant stellar contamination. Furthermore, stars are excluded from our
selection based on the $\chi^2_{opt+Y}$ measure. Even a cool dwarf star
with $H_{160} = 26$ mag shows $\chi^2_{opt+Y}\gtrsim100$ (see right panel
of Figure \ref{fig:colcol}). 

The low probability of stellar contamination is confirmed through SED
fitting. In addition to galaxy templates, we fit all four candidates with
stellar templates including observed dwarf spectra from
\citet{Burgasser04a}. None of these  fit any of our galaxy candidates
(likelihood for stellar contamination $<10^{-4}$), based purely on the
$HST$ photometry. We therefore conclude that it is very unlikely that any of our
candidates is a Galactic star.


\subsubsection{Photometric Scatter Simulations}
\label{sec:photscatter}

As we previously demonstrated \citep{Oesch13}, photometric scatter of
lower-redshift sources into the selection regions can be the most important source of
contamination for high-redshift LBG samples. This can be tested with
photometric scatter simulations based on real galaxies in the photometric
catalogs of fields where much deeper data are available in the same filters
as for the CANDELS data. In particular, we make use of the XDF and the
HUDF09-2 datasets, which have $HST$ optical (and NIR) data in the same
filters but have limits that are up to 2.5 mag fainter. 

From the deeper fields, we selected sources in the magnitude range of the
GOODS-N $z\sim10$ galaxy candidates, i.e., $H_{160} = 26-27.5$ mag, and we
applied photometric scatter as measured for real sources at those magnitudes
in the CANDELS data. 
In detail, we computed the average flux uncertainty of real sources in
our CANDELS catalogs and used these to add a Gaussian perturbation to the
flux measurements from the deep data.


The contamination fraction can then be estimated
based on applying the LBG selection criteria to this simulated catalogs, correcting for the GOODS-N
CANDELS survey area, and repeating this many times.
The resulting average number of contaminants per realization
is only 0.19.  Not unexpectedly, these contaminants are all found closer to the magnitude
limit at $H_{160}>27.0$ mag rather than at $H_{160}<27.0$ where the GOODS-N
$z\sim9-10$ candidates lie. This indicates that photometric scatter is not
significantly contaminating our sample at the observed brightness of these
candidates.

An independent estimate of the contamination fraction in the sample
can also be obtained from the best low-redshift SED fits by estimating the
probability with which such galaxy SEDs would be selected in the CANDELS
data. We therefore applied photometric scatter to the expected magnitudes of
the low-redshift SEDs shown in Figure \ref{fig:sedFits} and then applied our LBG
selection criteria. Repeating this $10^{6}$ times for all four candidates
results in an average contamination fraction of only 0.15 source per
realization, consistent with the 0.19 estimated above from the scatter
simulations based on the XDF and HUDF09-2 fields.

To summarize, the high quality, multiwavelength HST data and our
use of all information in the optical data based on the $\chi^2_{opt+Y}$
measure allows us to conclude that photometric scatter of lower redshift galaxies with known SEDs is
unlikely (consistent with our analysis of the photometric redshift
likelihood functions). However, we can not rule out contamination by
unusual, extremely rare sources not represented in the simulation databases.
The simulations we carried out assumed sources that have SEDs or photometric
characteristics consistent with documented results.  While it is quite
unlikely that we have identified sources with very unusual SEDs, the
possibility remains, though finding four such undocumented sources seems a remote possibility.  Deep spectroscopy of these candidates is the only way
to ultimately establish their true nature.

\begin{figure}[tbp]
	\centering
\includegraphics[width=0.9\linewidth]{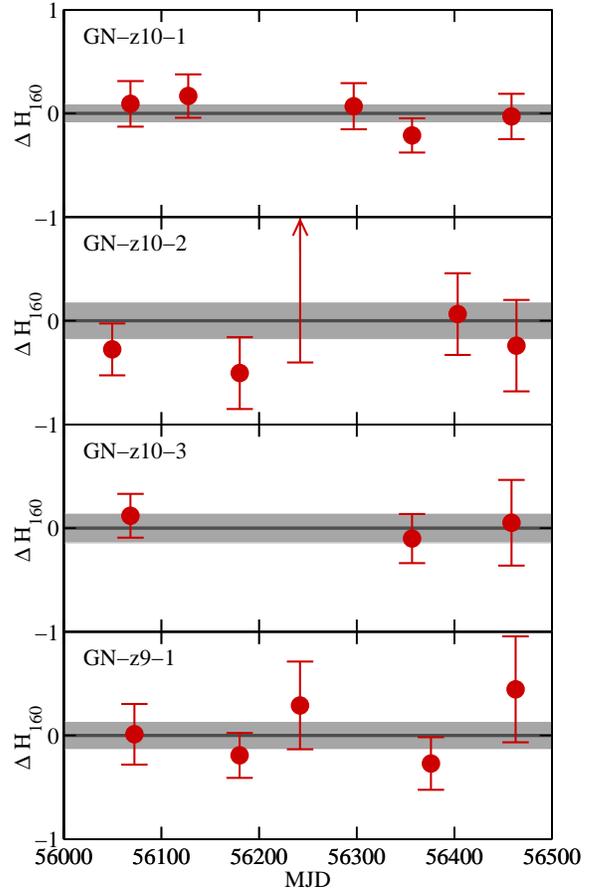}	  
  \caption{The variability of our four GOODS-N $z\sim9-10$ candidates in the $H_{160}$-band. The CANDELS GOODS-N $H_{160}$ exposures in the relevant regions were sorted by their acquisition time and binned in $\sim100$ day wide bins in order to obtain five independent $H_{160}$ images at different times. The bins were chosen as a compromise between having a significant number and having adequate S/N per bin. Given the data acquisition schedule, the times vary significantly from region to region.  The figure shows the magnitude differences with 1$\sigma$ errorbars for each source at the median acquisition time of each stack around that particular source as a function of the Modified Julian Date (MJD). Positive values correspond to fainter magnitude measurements. The gray region corresponds to the final magnitude uncertainty for each source. For almost all subsets a $>2\sigma$ detection is seen at the expected position. The exceptions occurred for the middle two sources. Source GN-z10-3 is in the CANDELS-Wide area and is thus covered by fewer exposures than the other sources. In two of our splits, the source landed on a masked part of the WFC3/IR detector and therefore could not be detected in those images. For GN-z10-2, we find no significant detection in the middle stack, where we show a $2\sigma$ upper limit on the source flux. Overall, the testing showed that there was no evidence for variability on timescales of 100 days at levels over a couple of tenths of a magnitude for any of these candidates. }
	\label{fig:variability}
\end{figure}

\subsection{AGN Contribution?}

As pointed out earlier, the GOODS-N $z\sim10$ galaxy candidates are
very compact and very bright. This raises the possibility that some of these
sources host an active galactic nucleus (AGN), which contributes or even
dominates the observed fluxes in the $H_{160}$-band. While it would be
surprising (though very interesting) to see significant AGN activity just a
few hundred million years after the formation of the first stars, without
spectroscopic observations, it is of course nearly impossible to reliably
assess such a contribution. However, our imaging data can provide some
first constraints. 

At least the two brighter sources in our sample appear to be resolved.
However, as is usually the case for compact high redshift sources, the
current data do not exclude a point source in the center of a more extended
star forming region. For such objects, probably the most stringent
constraint on an AGN contribution can  be obtained from a variability
analysis. AGN flux variations are seen over essentially all timescales and
would be a clear indicator for nuclear accretion activity \citep[for a
review see, e.g.,][]{Ulrich97}.  

Given that the GOODS-N CANDELS data were acquired
over almost a 2-yr timescale, we can directly test for variability. We have
therefore split the data into five separate subsplits, sorted by acquisition
time, and have reduced these frames separately. This was done for the
$H_{160}$-band images of each of our candidates. We then remeasured the
magnitudes of these sources via the dual image mode of SExtractor with the
full stack as the detection image. The resulting flux variation is shown in
Figure \ref{fig:variability}. 

For two of our sources, we do not detect a significant signal ($>2\sigma$)
at a couple of epochs. This results from the
low exposure times in those stacks when the source falls on masked regions
of the WFC3/IR detector. Thus the lack of data in those epochs is not
indicative of real variation.  

Evaluation of the measurements for each source shows that none of the
sources displays a statistically-significant variation.  This provides some
indicative evidence against AGN contributions in these candidates. Due to
the limited depth of the data, however, smaller amplitude flux variability
(of the order of a couple of tenths of a magnitude)
can not be ruled out.

\subsection{Possible Lensing Magnification}

Given the brightness of our candidates, it is interesting to ask whether
any of these could be significantly magnified by a foreground source. Even
though none of the candidates appears to be highly magnified (none have the
significant elongation which would be a clear sign for very high
magnification), smaller values of magnification are possible.
\citet{Wyithe11} estimated that very high-redshift luminosity functions
could be significantly distorted due to a magnification bias once the
characteristic magnitude lies $\sim2$ mag below the survey limit. This
could indeed be the case with our GOODS-N data.

We therefore examined the neighbors of all four candidates. The two
lowest-redshift ones do not show a very bright source nearby and are
therefore unlikely to be affected by lensing. However, the two
highest-redshift sources do show neighbors within 2\farcs9 and 1\farcs2
(see the top two rows in Figure \ref{fig:stampszgtr8}).  To see if
magnification was contributing to their unusual brightness we  
estimated their possible magnification bias based on the simplified
assumption of a Singular Isothermal Sphere (SIS) lens \citep[see,
e.g.,][]{Schneider06}.

\vspace{0.05cm} \hspace{0.2cm} $\bullet$ \textit{GN-z10-1:} Our
highest-redshift candidate shows a neighbor with $H_{160}= 24.7$ mag and
half-light radius $r_{1/2} = $0\farcs2 at a distance of  1\farcs2. 
Our photometric redshift analysis of this source indicates that it likely
lies at $z_{phot} = 1.8$, and has a stellar mass of $\log_{10}M=9.1 M_\odot$. While
we do not have any information on the velocity dispersion $\sigma_v$ of
this galaxy, we can obtain a rough estimate based on the virial theorem.
Assuming that all the mass is contained within 2$r_{1/2}$, we estimate
$\sigma_v = 27$ km/s.  This seems low and so we took a more conservative
assumption of a SIS with $\sigma_v = 50$ km/s. Even for this dispersion
this source has an Einstein radius for lensing a $z\sim10$ source of
$<0$\farcs04, resulting in a magnification of $<4$\% at the separation of
the candidate.

\vspace{0.05cm}
\hspace{0.2cm} $\bullet$ \textit{GN-z10-2:}
This candidate lies 2\farcs9 from a bright galaxy at a spectroscopic redshift $z_{spec} = 1.02$ \citep{Barger08}. The foreground source has $\log_{10} M = 10.8 M_\odot$ and half-light radius of $r_{1/2}=$0\farcs5, resulting in an estimate of $\sigma_v \sim 125$ km/s. A SIS model with these parameters has an Einstein radius of 0\farcs3 for lensing a $z\sim10$ galaxy resulting in a possible magnification of $11\%$ for this source. 

From the above considerations, we conclude that lensing magnification is
most likely not significant for our sample, amounting to at most $0.1$
mag.  Given the small magnifications and the uncertainties in the above
estimates, we do not correct for any possible magnification.

\section{Implications for the Galaxy Population at $z>8$}
\label{sec:results}

The detection of four very bright $z>9$ galaxy candidates in GOODS-N is
quite surprising given the dearth of candidates in the very similar
GOODS-S data as well as in the much deeper data in the three HUDF09
fields. In this section, we present the implications of these detections
for the interpretation of galaxy evolution at $z>8$.  
In particular, we will combine the GOODS-N data with our previous $z\sim10$ search over the GOODS-S and the HUDF09/XDF fields. In order to do so, we restrict our analysis to the three
sources in the GOODS-N sample that satisfy a more stringent $J_{125}-H_{160}>1.2$
criterion. This excludes only the $z\sim9$ candidate GN-z9-1, while the other three GOODS-N
candidates which have $z_{phot}=9.5-10.2$ are included.

Motivated by the discovery of the bright sources in GOODS-N we also systematically re-analyzed the CANDELS GOODS-S data with search criteria that are better matched to those used in GOODS-N. 
As discussed in detail in the appendix, we indeed identified two potential $z>9$ galaxy candidates which are also relatively bright ($H_{160}=26.6$ and 26.9). One of them (GS-z10-1) shows $J_{125}-H_{160}>1.2$ and has a photometric redshift of $z_{phot} = 9.9\pm0.5$. We will therefore also include this additional, new source in our subsequent analysis.

In section \ref{sec:SFRDevol}, we will determine the evolution of the cosmic SFRD to $z\sim10$, for which we include the two additional $z\sim10$ candidates from the CLASH survey \citep{Zheng12,Coe13}. Given the uncertainties in the lensing magnification we will not include these sources in our constraints on the UV LF, however.

\begin{figure}[tbp]
	\centering
	\includegraphics[width=\linewidth]{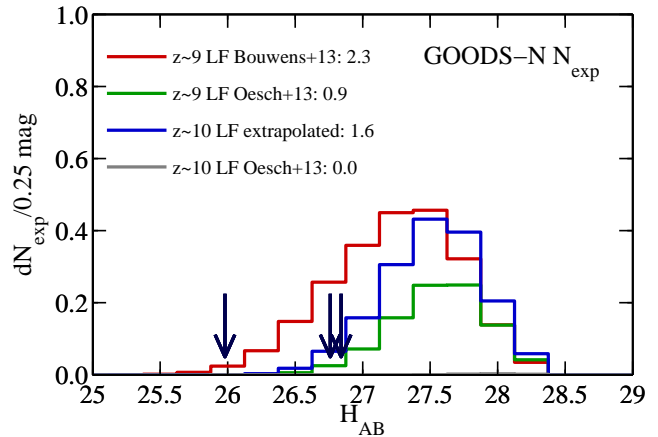}  
  \caption{Expected number of $z\sim10$ galaxy candidates per bin of 0.25 mag in the GOODS-N field for different assumptions about the UV LFs. These include previous estimates at $z\sim9$, $z\sim10$, and an extrapolation of the lower redshift UV LF trends to $z\sim10$. The total expected numbers for each assumption are indicated in the legend. The magnitudes of the three observed candidate $z\sim10$ galaxies in GOODS-N are indicated by black arrows. All the tested LFs produce a distribution which peaks at significantly fainter magnitudes than observed in the sample. Furthermore, all assumed LFs result in a lower number of total expected sources at the magnitudes of the three that are observed. In particular, not a single source would be expected to be seen in GOODS-N if the estimated UV LF from the HUDF09/12 and GOODS-S fields (Oesch et al. 2013) was correct. The detection of three bright $z\sim10$ candidates may thus be evidence for significant cosmic variance at the bright end of the UV LF.  Such bright sources have significant consequences on the best-fit UV LF at $z\sim10$.  }
	\label{fig:NexpGOODSNonly}
\end{figure}

\subsection{The Expected Abundance of $z\sim10$ Galaxies in GOODS-N}
\label{sec:NexpGOODSN}


In order to compute the expected number of sources for a given UV LF, we use extensive simulations of artificial galaxies inserted in the real data to estimate the selection volume. The artificial sources were detected and re-selected in the same manner as the original sources, from which we estimate the completeness $C(m)$ and selection probabilities $S(z,m)$ as a function of $H_{160}$ magnitude $m$ and redshift $z$. 
For more information on the simulation setup see \citet{Oesch13}.

These simulations allow us to statistically correct for the fact that our catalogs are missing a fraction of real high-redshift sources due to blending with foreground galaxies which amounts to a typical incompleteness of 20\% across these HST fields.

Given the selection function and the completeness, we can compute the number of expected sources in bins of magnitude for a given LF $\phi(M)$:

\begin{equation}
N^\mathrm{exp}(m)  = \int_{\Delta m} dm \int dz \frac{dV}{dz} S(m,z)C(m)\phi(M[m,z])
\end{equation}

Figure \ref{fig:NexpGOODSNonly} shows the results of this calculation for several assumed LFs. Most importantly, for the best-fit $z\sim10$ LF from the GOODS-S+HUDF09/12 \citep{Oesch13} not a single $z\sim10$ galaxy candidate was expected to be seen in GOODS-N. 

Additionally, we tested the expected number of sources based on a simple extrapolation of lower redshift LF trends to $z\sim10$. This extrapolation is based on the parametrization of \citet{Bouwens11c}, who measured the UV LF evolution across $z\sim4$ to $z\sim8$ and found: $\phi* = 1.14\times10^{-3} $Mpc$^{-3}$mag$^{-1} = $const,  $\alpha = -1.73 = $const and $M*(z) = -20.29 + 0.33 \times (z - 6)$. This extrapolation results in an assumed $M*(z=10) = -18.97$, and predicts 1.6 sources overall in GOODS-N, but only 0.14 at $H_{160}<27$ mag.


Very similar numbers are predicted by the theoretical $z\sim10$ UV LF model of \citet{Tacchella12}, from which one would have expected to see 1.9 sources overall in GOODS-N, of which only 0.5 were expected at $H_{160}<27$ mag.
Even for other test LFs, such as the $z\sim9$ UV LF estimates from
\citet{Oesch13} and \citet{Bouwens12CLASH}, only $\sim2$ sources were
expected to be seen in GOODS-N. However, they are expected to be fainter
with $H_{160}>27$ mag. The detection of three bright sources at
$H_{160}<27$ mag is therefore quite unexpected, given our previous
constraints on the UV LFs from multiple surveys. The three GOODS-N
$z\sim10$ candidates should therefore have interesting implications for
the LF at early times and also for the cosmic SFRD evolution at $z>8$.

It is instructive to see what must be done with current UV LFs to get as many as three bright $z\sim9-10$ galaxies.  For example, 3.4  bright  ($H_{160}<26.8$ mag) $z\sim9-10$ candidates are predicted in the GOODS-N data only if the UV LF at $z\sim9-10$ was the same as at $z\sim8$ \citep[using, e.g., the results of][]{McLure13}.
However, using the same unchanged $z\sim8$ LF predicts a total of 11 $z\sim10$ candidates in the GOODS-N data including fainter sources, and it predicts 42 sources in the whole combined search field (see section \ref{sec:combinedz10}).
This clearly is not the case and is securely ruled out.

\subsection{Discussion of Cosmic Variance}

As we outlined in previous
sections, we find no reason that the GOODS-N sample is heavily contaminated
by lower redshift sources. However, the detection of three bright candidates at $H_{160}<27$ mag was quite unexpected.  
Together with the non-detection of any source in the intermediate magnitude range of current
$z\sim10$ searches, this may indicate that $z\sim10$ galaxies are subject
to substantial cosmic variance. 


We used the publicly available cosmic variance
calculator\footnote{\url{http://casa.colorado.edu/\~{}trenti/CosmicVariance.html}}
of \citet{Trenti08} to estimate the likely impact of this on 
$z\sim10$ galaxy searches \citep[see also][]{Robertson10a}. Based on a simple halo
abundance matching, one expects a cosmic variance of $40-45\%$ per 4.7
arcmin$^2$ WFC3/IR pointing, depending on the assumptions about the halo
occupation fraction. For the field layout of the $\sim150$ arcmin$^2$
GOODS-N or GOODS-S WFC3/IR data, the expected cosmic variance ranges
between 15 to 20\%.  

Given the very low number of expected sources in
each survey, the variance is completely dominated by Poissonian statistics.
For instance, the chance of finding three or more $z\sim10$ galaxy candidates in the GOODS-N
field when 1.6 sources are expected is 22\%, independent of whether one
assumes a 20\% cosmic variance or not, on top of Poissonian statistics. 



What the analysis of the GOODS-N data  shows, is that larger datasets have
to be analyzed for reliable measurements of the UV LFs at very high
redshifts in order to overcome the limitations of Poissonian statistics.  It will therefore be interesting to explore the upcoming HST
Frontier Fields, which will add another 8 to 12 deep field pointings in
which one would expect $\sim0.5-1$ sources each for a (hopefully) much more
reliable sampling of the UV LF, particularly at intermediate magnitudes.


\subsection{The Combined $z\sim10$ Galaxy Sample from the GOODS-N+S and HUDF09/XDF}
\label{sec:combinedz10}

In order to constrain the cosmic SFRD evolution at $z>8$, we
combine our analysis of GOODS-N with previous $z\sim10$ galaxy
searches. In particular, we directly use the results from \citet{Oesch12b}
and \citet{Oesch13}, who analyzed all three ultra-deep HUDF09 fields
(including the new HUDF12 data in the XDF reduction) as well as the complete CANDELS GOODS-S
data. 
In these fields, only one viable $z\sim10$ candidate was previously
identified satisfying $J_{125}-H_{160}>1.2$. This source (XDFj-38126243) is
extremely faint with $H_{160}=29.8$.  It was found in the deepest WFC3/IR
imaging available in the HUDF/XDF field. 

We have systematically re-analyzed the CANDELS GOODS-S data and found one additional  bright  $z\sim10$ galaxy candidate. This source, GS-z10-1, together with a slightly lower redshift candidate is discussed in detail in the appendix.

\begin{figure}[tbp]
	\centering
	\includegraphics[width=\linewidth]{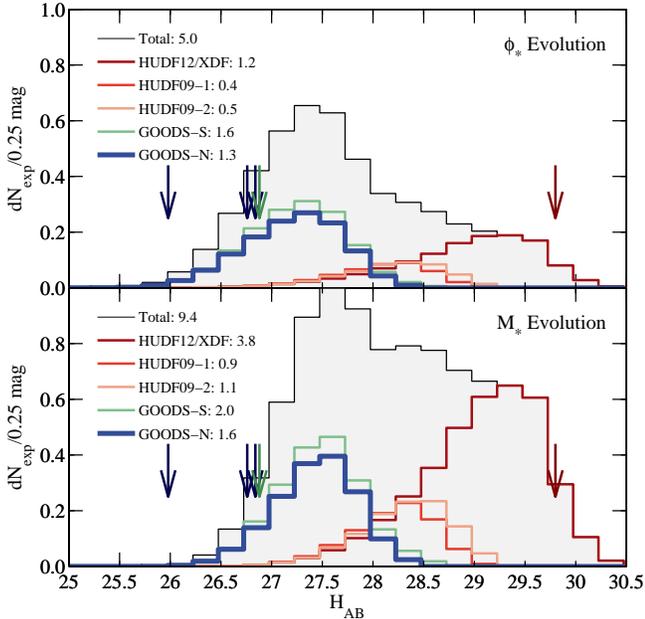}
  \caption{ Expected number of $z\sim10$ candidates per bin of 0.25 mag in the different fields used in our analysis for the best-fit UV LFs under the different assumptions of $\phi_*$-only (top) and $M_*$-only evolution (bottom). The different lines correspond to different survey fields as shown in the legend, while the shaded gray region corresponds to the total in our analysis. The legends list the breakdown of expected sources in each field for the two assumed LF evolutions.
The downward pointing arrows indicate the magnitudes of the three GOODS-N $z\sim10$ candidates (dark blue), the one GOODS-S source (green), and the XDF candidate (dark red). Interestingly, the detected candidates only cover the tails of the expected magnitude distribution, with no sources being detected around any of the peaks.
The two assumptions result in quite different magnitude distributions and expected number of candidates in fields of different depths, which has important consequences for planning future surveys for such high-redshift sources.
  }
	\label{fig:Nexp}
\end{figure}

The total $z\sim10$ galaxy sample with $J_{125}-H_{160}>1.2$ that is used in the reminder of this paper thus comprises three bright sources in GOODS-N, one bright candidate from GOODS-S, and one faint source from the XDF data. The most striking feature of this sample is that no sources are found at intermediate magnitudes at $H_{160}=27-29$ (see, e.g., Fig \ref{fig:Nexp}). This, along with the very small sample size, will make it very challenging to derive a reliable LF (see next section).


The selection functions and completeness curves for the GOODS-S and HUDF09 fields have
previously been computed by \citet{Oesch12a} and \citet{Oesch13}. The use of an identical approach to the GOODS-N analysis in this paper allows us to
directly combine all search fields for a total measurement of the galaxy
number density at $z\sim10$. Given our systematic re-analysis of the GOODS-S field, we have updated the completeness and selection functions to be consistent with our new SExtractor catalogs.


\begin{deluxetable*}{lccccccc}
\tablecaption{Summary of $z\sim10$ UV LF and SFRD Estimates \label{tab:lfsummary}}
\tablewidth{0 pt}
\tablecolumns{5}
\tablehead{   & $\log_{10}\phi_*$  [Mpc$^{-3}$mag$^{-1}$]  &  $M_{UV}^*$ [mag]  &  $\alpha$ & $N_{exp}^{tot}$\tablenotemark{*} & $\log_{10}\dot{\rho}_{*}$  [\msol~ yr$^{-1}$Mpc$^{-3}$] \tablenotemark{$\dagger$}}

\startdata
This Work (from candidates)  & \nodata   &  \nodata &  \nodata & \nodata & $-3.25\pm0.35$ \\
This Work  ($\phi_*$ evolution)				  & $-4.27\pm0.21$ & $-20.12$ (fixed)  & $-2.02$ (fixed) &   $5.0^{+3.4}_{-2.2}$ & $-3.22\pm0.21$\\
This Work ($M_*$ evolution) 				        & $-3.35$ (fixed)  & $-19.36\pm0.15$  & $-2.02$ (fixed) &  $9.4^{+4.2}_{-3.0}$  &  $-2.80\pm0.11$ \\

\noalign{\vskip .7ex} \hline \noalign{\vskip 1ex}
\citet{Oesch13} 				  & $-2.94$ (fixed)  & $-17.7\pm0.7$  & $-1.73$ (fixed)  & $0.6^{+2.5}_{-0.5}$   & $-3.7^{+0.7}_{-0.9}$ 
%

\enddata
\tablenotetext{$\dagger$}{The SFRD measurement is limited at $M_{UV}<-17.7$, the luminosity limit of the HUDF12/XDF data.}
\tablenotetext{*}{Total number of $z\sim10$ candidates with $J_{125}-H_{160}>1.2$ expected to be seen in all the search fields of this paper. These include GOODS-North, GOODS-South, the HUDF09 parallel fields as well as the HUDF12/XDF field, in which we identified a total of five candidate galaxies.}
\end{deluxetable*}

\begin{figure}[tbp]
	\centering
	\includegraphics[width=\linewidth]{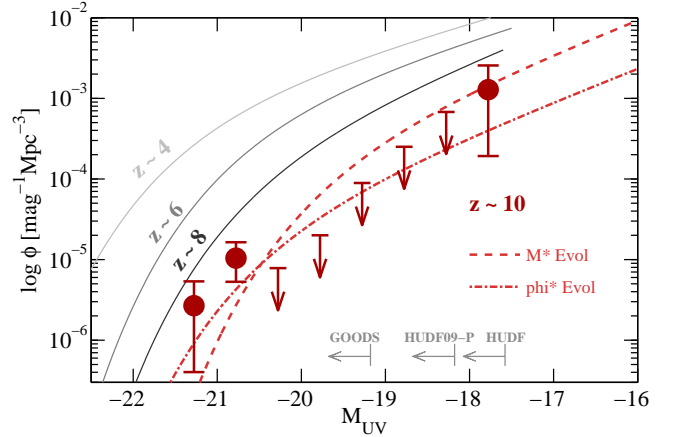}  
  \caption{Improved constraints on the $z\sim10$ UV LF from the combined $z\sim10$ search using the blank-field GOODS-N, GOODS-S, and HUDF09/12/XDF WFC3/IR datasets. The additional data from fields other than the GOODS-N are taken directly from \citet{Oesch13}. The dark red circles indicate the step-wise UV LF estimates in bins of 0.5 mag using the four GOODS-N+S and the one HUDF12/XDF $z\sim10$ galaxy candidates satisfying $J_{125}-H_{160}>1.2$. Upper limits are $1\sigma$. The dashed line represents the best-fit $M_*$-only evolution relative to the $z\sim8$ UV LF, while the dot-dashed line shows the same for $\phi_*$-only evolution.
   Lower redshift LFs are shown as gray
solid lines for illustration of the LF evolution trends \citep{Bouwens07,Bouwens12b,McLure13}. Evolution in $\phi_*$ appears to better match the full dataset including the new results from GOODS-N.  We have not included the CLASH survey
 candidates \citep{Zheng12,Coe13} given the uncertainties in the lensing magnification.
 }
	\label{fig:LFevol}
\end{figure}

\subsection{Improved Constraints on the UV Luminosity Function at $z\sim10$}
\label{sec:z10UVLF}

The dearth of $z\sim10$ candidate sources in the intermediate magnitude
range $H_{160}=27-29$ mag (see Figure \ref{fig:Nexp}), makes it challenging
to provide a meaningful Schechter LF fit \citep{Schechter76} to the
observed sources. A simple power-law might provide a better description of
the UV LF at such high redshifts. However, the widespread use of Schechter
LF fit at lower redshifts $z\sim4-8$ and in previous papers at $z\sim9-10$
suggests that use of the same formalism at $z\sim10$ is useful for
comparative purposes. Furthermore, theoretical models and simulations still point toward a Schechter-like function \citep[e.g.]{Trenti10,Lacey11,Tacchella12}. 
We thus update our previous estimates of the
Schechter function parameters based on the combined dataset of
GOODS-N/S+HUDF09/HUDF12/XDF.

In our previous analysis we assumed the characteristic luminosity, $M_*$,
to be the main parameter of the Schechter function to evolve to higher
redshift. This was motivated by previous $z\sim4-8$ measurements of the UV
LF. However, this assumption is called into question with the three
detections in GOODS-N and the one bright source in GOODS-S, because, as we shall see below, an $M_*$-only evolution results in a substantial over-prediction of the total number of candidates in our search fields.

In order to show this, we first determine our baseline lower redshift UV LF
model, relative to which we will measure the evolutionary trends. Over the
last few years, several $z\sim8$ UV LF determinations have been published
by several teams based on WFC3/IR datasets
\citep[e.g.,][]{Bouwens10a,Bouwens11c,Yan11,Bradley12,Oesch12b,Lorenzoni13,Schenker13,McLure13}.
The most recent determinations among these that use several search fields are
all in  good agreement with each other, and returned consistent estimates
of the $z\sim8$ UV LF Schechter function parameters. As a baseline model we
adopt the values from \citet{McLure13} which represents the widest area
study to date and its UV LF parameters represent a good average of recent
results from several teams \citep[see, e.g., Table 6 of][]{Schenker13}.

Hence, for the $z\sim8$ baseline, we adopt $\log_{10}\phi_*(z=8) = -3.35$
Mpc$^{-3}$mag$^{-1}$, $M_*(z=8) = -20.12$ mag, and $\alpha(z=8) = -2.02$
\citep{McLure13}.  We then estimate the $z\sim10$ UV LF parameters relative
to this baseline model by varying one parameter at a time. In particular,
we test for $M_*$- and $\phi_*$-evolution.

\begin{deluxetable}{ccc}
\tablecaption{Stepwise $z\sim10$ UV LF Based on the Full Dataset \label{tab:10lf}}
\tablewidth{0 pt}
\tablecolumns{2}
\tablehead{$M_{UV}$ [mag] & $\phi_*$  [10$^{-3}$Mpc$^{-3}$mag$^{-1}$]   }

\startdata

$-21.28$  &  $0.0027^{+0.0027}_{-0.0023}$  \\ 
$-20.78$  &  $0.010^{+0.006}_{-0.005}$  \\ 
$-20.28$  &  $<0.0078$  \\ 
$-19.78$  &  $<0.020$  \\ 
$-19.28$  &  $<0.089$  \\ 
$-18.78$  &  $<0.25$  \\ 
$-18.28$  &  $<0.68$  \\ 
$-17.78$  &  $1.3^{+1.3}_{-1.1}$  

\enddata

\tablecomments{Limits are 1$\sigma$ for a non-detection.}

\end{deluxetable}

The best-fit parameters were determined by minimizing the Poissonian
likelihood of observing $N_{obs}$ sources in a given magnitude bin when
$N_{exp}$ are expected from the LF: $\cal{L} $ $= \prod_{j} \prod_i
P(N^{\rm obs}_{j,i},N^{\rm exp}_{j,i})$, where $j$ runs over all fields,
$i$ runs over the magnitude bins of width 0.5 mag, and $P$ is the
Poissonian probability. 

Doing so for $M_*$-only evolution relative to the baseline model results in
a best-fit estimate of $M_*(z=10) = -19.36\pm0.15$. The expected magnitude
distribution of $z\sim10$ candidates for this LF is shown in the lower
panel of Figure \ref{fig:Nexp}, and the LF itself is shown as the dashed
line in Figure \ref{fig:LFevol}. This determination lies significantly
above the upper limits at intermediate magnitudes. With
9.4$^{+4.2}_{-3.0}$, the total expected number of $z\sim10$ galaxy
candidates from this LF is also larger than the five observed sources, though the difference is not very significant. A larger disagreement arises since these nine sources would all be expected at magnitudes fainter than
$H_{160}\sim27$ mag in the different search fields (Fig. \ref{fig:Nexp}). Yet these are not seen.

A somewhat more consistent result is achieved from a fit with $\phi_*$ evolution.
Assuming again the baseline $z\sim8$ LF parameters and varying only
$\phi_*$, we find a best-fit $\log_{10}\phi_*= -4.27\pm0.21$
Mpc$^{-3}$mag$^{-1}$, almost  an order of magnitude lower than the
$z\sim8$ normalization. This LF is shown in Figure \ref{fig:LFevol}. As can
be seen, it represents a better compromise between the detections at the
bright and faint end and the upper limits at intermediate luminosities. The
total expected number of sources for this model is $5.0^{+3.4}_{-2.2}$,
consistent with the observed number of sources, but again with a magnitude
distribution which peaks at $H_{160} = 27$-29 mag, where we do not detect any
candidates (see Figure \ref{fig:Nexp}). 
These best-fit UV LF parameters and the corresponding total number of
expected sources are summarized in Table \ref{tab:lfsummary}, and the
stepwise $z\sim10$ UV LF constraints are tabulated in Table \ref{tab:10lf}.

While previous results in the literature  suggested the main evolving parameter of 
the UV LF at $z\sim4-8$ to be predominantly the characteristic luminosity $L^*$, evolution that is
dominantly $\phi^*$  is consistent within the uncertainties. In fact, evolution in the normalization of the UV LF may
be more easily accommodated by current theoretical models.   
As discussed in \S5.5 of \citet{Bouwens08}, one challenge with $L^*$
evolution being the dominant form of evolution in the LF is that it
requires some physical mechanism to impose a cut-off at a specific
luminosity (and likely mass) in the $UV$ LF and for that luminosity to
depend on redshift.  Since it is not clear what physical process would
cause the cut-off to depend on redshift, simulators often find very
little evolution in $L^*$ \citep[e.g.,][]{Jaacks12a}.

Even for the best-fit LF from $\phi_*$-evolution, we only expect to see 0.4 sources brighter than $H_{160} = 27$ mag in GOODS-N, and 0.9 in all fields combined. Detecting three such bright sources in the GOODS-N alone is very unlikely, with a probability that is only 0.8\% from Poissonian statistics. Unless these sources are found to be at low redshift (which appears unlikely) the detection of so many bright galaxies in GOODS-N would indicate the need for much larger cosmic variance than predicted by a simple halo abundance matching. This could be caused, e.g., by bursty and highly biased star-formation at very high redshift with low duty cycle \citep[see, e.g.,][]{Jaacks12b,Wyithe13}.

Alternatively, given our somewhat improbable mix of $z\sim10$
detections and limits, i.e., with our $z\sim10$ candidate galaxies
only being found at the extrema of the luminosity range probed (see
Figure \ref{fig:LFevol}), one other possibility one could consider is
a non-Schechter-like form for the LF at $z>6$, as has already been
speculated elsewhere \citep[e.g.,][]{Bouwens11c,Bowler12}.

\begin{figure*}[tbp]
	\centering
	\includegraphics[width=0.72\linewidth]{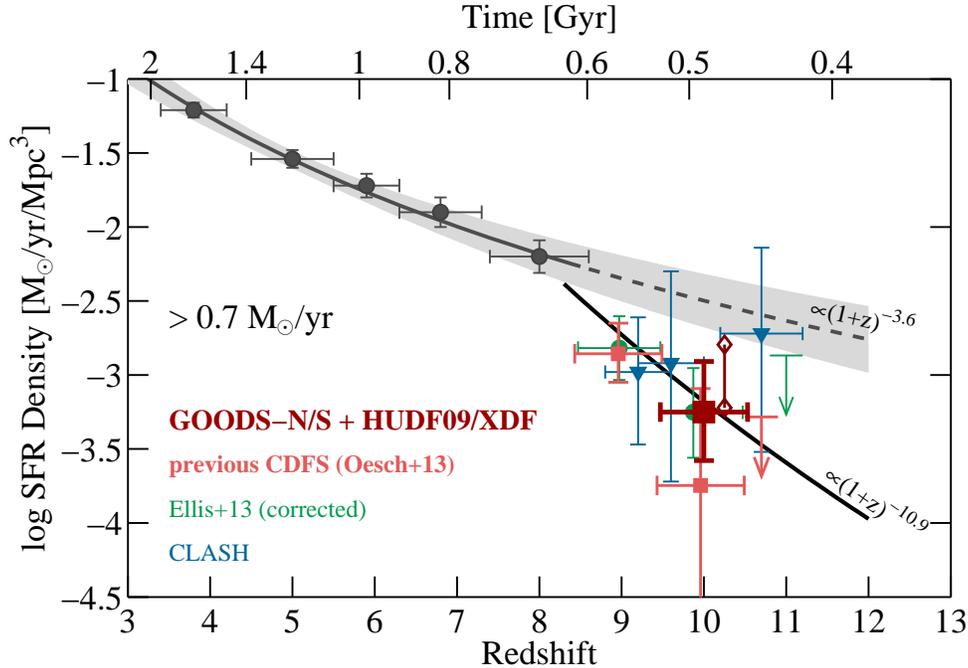}
  \caption{ The redshift evolution of the cosmic star-formation rate density (SFRD)  $\dot\rho_*$ above
 a star-formation limit $>0.7 M_\odot$yr$^{-1}$ including the new GOODS-N $z\sim10$ galaxy candidates. 
 The lower redshift SFRD estimates are based on LBG UV LFs
from \citet{Bouwens07,Bouwens12b} including dust corrections. The gray band represents their $1\sigma$ uncertainty.
The new measurement from the five detected candidates in the combined CANDELS GOODS-N/S and the HUDF09/12/XDF dataset is shown as the dark red square. The individual SFRD with errorbars were computed from the UV LD of the individually detected sources. Open diamonds connected with a vertical line represent the SFRDs as estimated based on integrating the best-fit UV LFs down to the corresponding luminosity limit of $M_{UV}=-17.7$ (see Table \ref{tab:lfsummary}). The upper diamond represents $M_*$ evolution, and the lower diamond is derived from $\phi_*$ evolution. These estimates are offset to $z=10.25$ for clarity.
Previous measurements of the SFRD at $z>8$ are shown from a combination of HUDF09/12+GOODS-S \cite[pale red;][]{Oesch13}
 as well as from CLASH cluster detections \citep[blue triangles;][]{Bouwens12CLASH,Coe13,Zheng12}. Additionally, we also show the results of the HUDF12 field only from \citet{Ellis13} (green circles). We corrected down their $z\sim10$ point by a factor $2\times$ to account for our removal of a source that was shown to be a diffraction spike \citep[see][]{Oesch13}.
 When combining all the measurements of the SFRD at $z\geq8$ from different fields we find $\log \dot\rho_* \propto (1+z)^{-10.9\pm2.5}$ (black solid line), significantly steeper than the lower redshift trends which only fall off as $(1+z)^{-3.6}$ (gray line). The current data at $z>8$ show that the cosmic SFRD is very likely to be increasing dramatically, by roughly an order of magnitude, in the 170 Myr from $z\sim10$ to $z\sim8$.
   }
	\label{fig:SFRDevol}
\end{figure*}

%
%
%
%
%

\subsection{The Evolution of the Cosmic SFRD at $z>8$}
\label{sec:SFRDevol}

The most recent WFC3/IR datasets from several independent surveys have
enabled determinations of the cosmic SFRD at redshifts $z\sim9$ and
$z\sim10$ that were thought to be largely inaccessible for quantitative
constraints on LFs or the SFRD. Estimates of the SFRD have been performed
independently from a small sample of sources from the CLASH program
\citep{Bouwens12CLASH,Zheng12,Coe13} and from the HUDF09 and HUDF12 surveys
\citep{Bouwens11c,Oesch12a,Oesch13,Ellis13,McLure13}. While the conclusions
from these separate analyses disagree on the SFRD evolutionary trends,
\citet{Oesch13} have shown that combining all the results from the
literature, the SFRD appears to increase by an order of magnitude in just 170 Myr from
$z\sim10$ to $z\sim8$, down to the current detection limits of the
HUDF09/12/XDF dataset.

Here we extend our previous analysis from the CANDELS GOODS-S and
HUDF09/12/XDF for an updated estimate of the cosmic SFRD at $z\sim10$ with
the inclusion of the new sources found in the GOODS-N field, and in the GOODS-S. Figure \ref{fig:SFRDevol} shows the new
results plus other measurements at $z>8$ from the literature and the
cosmic SFRD evolution constrained by these data. 

The SFRD was computed directly from the observed UV luminosity density (LD) of the three GOODS-N, the single bright GOODS-S,
and the one XDF $z\sim10$ galaxy candidates. The LDs were converted to a SFRD
using the conversion factor of \citet{Madau98}, assuming a Salpeter initial
mass function\footnote{SFR($M_\odot$ yr$^{-1}$) = 1.4 $\times$ 10$^{-28}$
L$_{1500}$ (erg s$^{-1}$ Hz$^{-1}$) \citep{Kennicutt98} }. Since the
HUDF12/XDF data reaches down to $M_{UV} = -17.7$ mag, the derived SFRD is
limited at SFR$ > 0.7 M_\odot$yr$^{-1}$.  For the $z>8$ points, we did not
perform any dust correction, because it is expected to be negligible based
on the evolution of the UV continuum slope distribution at lower redshift
\citep[e.g.,][]{Bouwens12a,Bouwens13,Dunlop13,Finkelstein12,Wilkins11b}.

The direct SFRD from the five observed candidates is
$\log_{10}\dot{\rho}_{*} = -3.25\pm0.35$  \msol~yr$^{-1}$Mpc$^{-3}$. As can
be seen from the summary in Table \ref{tab:lfsummary}, this is  a factor $0.45$ dex higher compared to our previous estimate using only the one HUDF12/XDF candidate. 
However, it remains quite consistent with our previous estimate that a very large change occurs in the SFRD in the 170 Myr from $z\sim10$ to $z\sim8$.  With the new data the
build-up remains strong at $1.05\pm0.38$ dex from $z\sim10$ to $z\sim8$, i.e., by an order
of magnitude.

Together with the direct SFRD as measured from the five detected sources in
the XDF and GOODS-N+S, Figure \ref{fig:SFRDevol} also shows the SFR densities
of the two best-fit UV LFs we derived in the previous Section (for a
summary see also Table \ref{tab:lfsummary}). In particular, the best-fit
$M_*$ evolution results in a significantly higher SFRD, essentially equal
to the current $z\sim9$ estimates. However, we stress again that $M_*$
evolution should have resulted in nine detected $z\sim10$ candidates in
our search fields and should therefore be considered an upper limit.

Combining our updated SFRD estimate with previous analyses from
different datasets in the literature at $z>9$
\citep{Bouwens12CLASH,Coe13,Zheng12} the new best-fit evolution of
the cosmic SFRD at $z\geq8$ is $\log\dot\rho_* \propto
(1+z)^{10.9\pm2.5}$, which is almost unchanged from our previous
determination without the new luminous sources in GOODS-N and GOODS-S
\citep[$(1+z)^{11.4\pm3.1}$;][]{Oesch13b}. The small change is mostly due to
the fact that our new, combined SFRD measurement from all the CANDELS-Deep
and HUDF09/XDF data almost exactly falls on the previously estimated trend and that the LF is so steep that the integrated flux is dominated still by the lower luminosity sources.

\section{Robust Rest-Frame Optical Detections of $z\sim10$ Galaxies: Nebular Emission Lines and Stellar Masses}
\label{sec:IRACDet}

The most important result from our IRAC analysis in the previous sections is
that for $all$ three sources in GOODS-N for which the neighbor subtraction was
successful, we  detect a significant signal in at least one IRAC band. The
detection significances in the 4.5\,$\mu$m channel are $5.8\sigma$
(GN-z10-1), $4.5\sigma$ (GN-z10-3), and $6.2\sigma$
(GN-z9-1). Furthermore, the brightest of our candidates`
(GN-z10-1) is detected at $6.9\sigma$ also in the 3.6\,$\mu$m band of
the IRAC data (see also Figure \ref{fig:stampszgtr8}).
For this source, we therefore have two secure IRAC flux measurements in
addition to its $H_{160}$-detection. These measurements allow us to place
significant constraints on the SED of  GN-z10-1 and derive relatively
robust physical properties based on SED fitting.
Additionally, the lower redshift source in GOODS-S (GS-z9-1) is  detected in both IRAC channels, while only a faint 4.5~$\mu$m detection is found for the other GOODS-S $z\sim10$ candidate GS-z10-1 (see Figure \ref{fig:stamps_newcandidates} and Table \ref{tab:fluxNewCand} in the appendix).

Note that the Spitzer 4.5~$\mu m$ channel only probes up to rest-frame B-band at the very high redshift of our targets. The ability to identify an underlying old population of stars is therefore limited with these data. However, this is unlikely to be a significant issue for these galaxies given the young age of the Universe at $z\sim9-10$.

The most striking feature of the SED of GN-z10-1 is that it does not
show any Balmer break, having an IRAC color of $[3.6]-[4.5] =
-0.14\pm0.29$. Overall, the SED of this source is extremely flat with few
features. Using a constant star-formation history and a \citet{Bruzual03}
model with 0.5$Z_\odot$, the SED is best fit with the minimum allowed age
of only 10 Myr and a stellar mass of 10$^{8.7\pm0.3}$ \msol. Interestingly, the
SED-based SFR for this source is 55 $M_\odot/$yr, which is significantly
higher than the observed SFR based on its uncorrected UV luminosity and
standard conversion factor (this gives 14 $M_\odot/$yr). The high SED-derived SFR can be attributed largely to the 
non-negligible dust extinction required in the SED fit of $A_{UV} = 1.1$
mag, together with the young SED age.

\begin{figure}[tbp]
	\centering
	\includegraphics[width=0.99\linewidth]{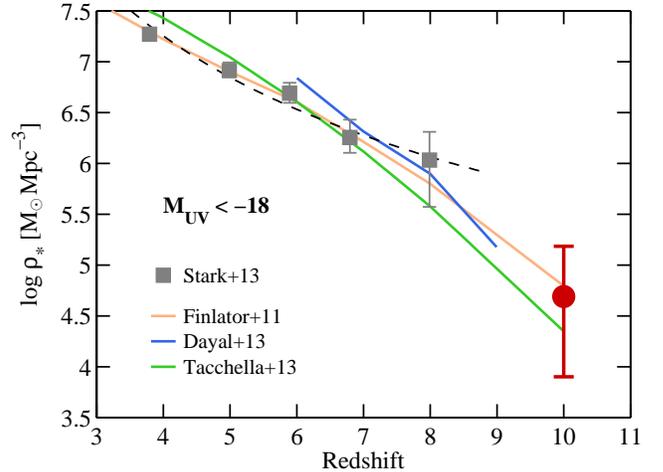}   
  \caption{ The redshift evolution of the cosmic stellar mass density in LBGs at $M_{UV}\lesssim-18$ to $z\sim10$. Gray squares represent the most recent determination of \citet{Stark13}, who used the measurements of \citet{Gonzalez11} and corrected these for the likely biases due to strong emission lines in the IRAC photometry for $z\sim5-8$ sources. Their empirical trends are indicated as a dashed black line. The dark red point is our first estimate of the $z\sim10$ stellar mass density based on the three GOODS-N $z=9.5-10.2$ galaxies, the one source in GOODS-S, and the XDF candidate. For the latter, we have assumed the same $M_*/L_{UV}$ ratio as for the average of the GOODS-N+S sources, for which these measurements can be done directly given the IRAC detections.  The stellar mass density is found to be more than an order of magnitude lower at $z\sim10$ than at $z\sim8$, consistent, as would be expected, with the drop in the cosmic SFRD down to $M_{UV}<-18$. This decrease to higher redshift is also consistent with the latest simulations and models, which are overplotted as colored lines \citep{Finlator11a,Dayal13,Tacchella12}.
  The current uncertainties on the stellar mass density measurements at $z\geq7$ are still significant, however. Deeper IRAC data over larger high-redshift samples would be ideal for reducing the uncertainties in the near future.
  }
	\label{fig:SMDevol}
\end{figure}

\begin{deluxetable}{lcccc}
\tablecaption{Physical Properties of GOODS-North and South $z\sim9-10$ Candidates\label{tab:masses}}
\tablewidth{\linewidth}
\tablecolumns{4}

\tablehead{ Source & $M_{UV}$ & $\log M_* [M_\odot]$ & $\log SSFR [\mathrm{yr}^{-1}]$ } 
\startdata 

GN-z10-1  & $-21.6 \pm 0.1$ & $8.7 \pm 0.3$ & $-7.0 \pm 0.6 $ \\ 
 GN-z10-2  & $-20.7 \pm 0.1$ & $(7.9)^\dagger$ & $(-7.1)^\dagger$ \\ 
 GN-z10-3  & $-20.6 \pm 0.2$ & $9.2 \pm 0.3$ & $-8.5 \pm 0.8 $ \\ 
 GN-z9-1  & $-20.7 \pm 0.1$ & $9.4 \pm 0.3$ & $-8.5 \pm 0.9 $  \\
 
 \noalign{\vskip .7ex} \hline \noalign{\vskip 1ex}

GS-z10-1 & $-20.6 \pm 0.2$ & $8.5 \pm 0.4$ & $-7.5 \pm 0.6 $ \\  
GS-z9-1  & $-20.8 \pm 0.2$ & $9.5 \pm 0.3$ & $-8.3 \pm 0.5 $ 
 
 \enddata 
 
 \tablenotetext{$\dagger$}{The stellar population parameters for source GN-z10-2 are highly uncertain due to uncertainties in the IRAC neighbor subtraction.}
 
\end{deluxetable}

Given that GN-z10-1 lies at $z_{phot}\sim10.2$, we can use its
$H_{160}-[3.6]$ color to estimate its UV continuum slope
$\beta$\footnote{defined as $f_\lambda \propto \lambda^\beta$}. This
results in $\beta=-2.0\pm0.2$, which is consistent with the UV continuum
slope of its best-fit SED ($\beta=-2.09$).  This may appear relatively red
for such a young high-redshift source. However, it is consistent with the
UV continuum slope evolutionary trends with luminosity found by \citet{Bouwens13}, given the
high luminosity of this source.

The SED fitting of the remaining two sources with clean IRAC photometry
gives stellar masses of 10$^{9.2\pm0.3}$ \msol~ for GN-z10-3 and 10$^{9.4\pm0.3}$ \msol\ for GN-z9-1 with SED ages of
$\sim300$ Myr. 
Similarly, the GOODS-S sources show stellar masses of 10$^{8.5\pm0.4}$ \msol~ and 10$^{9.5\pm0.3}$ \msol.
The typical stellar mass for these bright $z\sim9-10$ sources
therefore appears to be 10$^{9.0}$ \msol. The mass estimates, together with
estimates of the specific SFR (SSFR) of these sources are listed in Table 
\ref{tab:masses}. The stellar mass estimate of source  GN-z10-2 is highly
uncertain, because of the lack of clean IRAC photometry.


As can be seen from Table \ref{tab:masses}, the SSFRs of these sources are not very well constrained,
given the current photometry and redshift uncertainties. However, they all
lie higher than 2 Gyr$^{-1}$, where a possible flattening of the SSFR had been
suggested at $z\sim3-6$ \citep{Stark09,Gonzalez10}, contrary to
expectations from models. These galaxies
therefore provide some tentative evidence for a continued increase in the
SSFRs at higher redshifts \citep[see
also][]{Smit13,Stark13,Gonzalez12b,deBarros12,Schaerer10}.

Emission lines, which can significantly contaminate the IRAC photometry,
are implicitly accounted for in our SED library as discussed in Section
\ref{sec:SEDanalysis}. At the redshifts of our candidates ($z=9.2-10.2$),
the only significant line emission in the IRAC bands would be expected from
[\ion{O}{2}] or weaker Balmer lines. While these lines are likely not as
strong as [\ion{O}{3}]$_{4959,5007}$ or $H\alpha$, which have been
identified as significant contributors to IRAC fluxes of high redshift
galaxies at $z\sim4-8$
\citep[e.g.,][]{Schaerer09,Shim11,Labbe12,deBarros12,Stark13}, their
combined equivalent width can still be  substantial at very young ages
$10-50$Myr. In the most extreme cases it could be enough to boost the IRAC
photometry by up to $0.5$ magnitude, depending on the exact implementation
of the emission lines in the models and on poorly constrained quantities
such as metallicity, ionization parameter, reddening of the nebular
regions, and escape fraction.

While some uncertainties remain, these mass estimates allow us to derive a first estimate of
the stellar mass density at $z\sim10$ using the four bright $z>9.5$ sources from GOODS-N and GOODS-S. 
Using the volume density for these sources which we derived above, we estimate a mass
density for such galaxies of
$\log_{10}\rho_{*,z\sim10}(M_{UV}<-20.5) = 4.0^{+0.5}_{-0.6}$ \msol
Mpc$^{-3}$.  

To compare this with recent measurements in the literature at
lower redshifts, which include contributions to the stellar mass from sources with $M_{UV}<-18$ \citep[e.g.,][]{Gonzalez11,Stark13},
we have to fold in the contribution of fainter sources at $z\sim10$. This is quite
uncertain for several reasons: (1) we only have one faint candidate in our
combined dataset, and (2) this XDF source is too faint to be detected by IRAC,
even in the existing ultra-deep IUDF10 dataset over the XDF
\citep[see][]{Labbe12,Oesch13b}.

In order to get a first, simple estimate of the $z\sim10$ stellar mass
density to $M_{UV}<-18$, we thus assume that the XDF candidate has the same
$M_*/L_{UV}$ relation as the average bright GOODS-N+S galaxy, which are one order of
magnitude more UV-luminous. With this assumption, we derive
$\log_{10}\rho_{*,z\sim10}(M_{UV}<-18) = 4.7^{+0.5}_{-0.8}$ \msol
Mpc$^{-3}$.

Figure \ref{fig:SMDevol} shows that the estimated value of this stellar
mass density lies almost an order of magnitude below the extrapolated empirical
trends from recent determinations \citep{Gonzalez11,Labbe12,Stark13}. The large drop is mostly a consequence
of the drop in the UV LD from $z\sim8$ to $z\sim10$, given that the stellar
mass densities are all derived down to a fixed UV luminosity. The magnitude of the drop  is still
very uncertain given the current uncertainties on the stellar mass
density measurements at $z\geq7$.  
However, it is in overall very good agreement with theoretical model predictions as shown in the figure \citep{Finlator11a,Dayal13,Tacchella12}.
Larger $z>8$
galaxy samples with direct IRAC detections are required to derive more
robust estimates of the stellar mass densities in the future in order to be able to discriminate between different models.

Our analysis demonstrates the power of $Spitzer$/IRAC to probe galaxy
masses as early as 500 Myr after the Big Bang and indicates the potential
for further progress in the near future with a dedicated IRAC
survey.  

\section{Discussion and Conclusions}
\label{sec:summary}

We discuss the discovery of four very bright $z\sim9-10$ galaxy candidates
identified in the complete CANDELS GOODS-N dataset. These sources have
magnitudes in the range $H_{160} = 26-27$ mag, comparable to the two highly
magnified $z\sim10$ galaxy candidates found in the CLASH cluster dataset
\citep{Zheng12,Coe13}. However, these four GOODS-N candidates do not show
signs of significant magnification ($\lesssim0.1$ mag), and they are thus by far the most
luminous $z\sim10$ galaxy candidates detected with $HST$ to date.

Furthermore, these sources result in the first $>5\sigma$ $Spitzer$/IRAC
detections of $z>9$ galaxy candidates.  The brightest candidate is securely
detected at $6.9\sigma$ in the  3.6\,$\mu$m band and $5.8\sigma$ in the  4.5\,$\mu$m band of the 50 hr
$Spitzer$/IRAC data from the combined GOODS, SEDS, and S-CANDELS data. For
two of the three remaining candidates, those for which bright neighbor
subtraction was successful in the IRAC bands, we additionally detect
significant signal in the 4.5\,$\mu$m band at $4.5\sigma$ and $6.2\sigma$.
$Spitzer$/IRAC has thus been able to clearly detect rest-frame optical light out to $\sim$500 Myr
after the Big Bang.

Motivated by the bright galaxies in GOODS-N, we systematically re-analyzed the GOODS-S dataset for bright sources using similar selection criteria as in GOODS-N ($J_{125}-H_{160} > 0.5$) and updated SExtractor detection parameters. This analysis resulted in the detection of two similarly luminous sources in GOODS-S, both of which were also detected in Spitzer/IRAC, though one had marginal $2\sigma$ detections in both IRAC bands. These sources are discussed in the appendix.

The $Spitzer$/IRAC flux measurements, together with the extensive deep
$HST$ data, allow us to constrain any lower-redshift
contamination.  Extensive tests were done (see Section 3) to evaluate the
likelihood that these sources could be other than high-redshift $z\sim9-10$
galaxies.  The results suggest that the most plausible outcome is that
these galaxy candidates are really at $z\sim9-10$.  Yet we cannot rule out
that they constitute very unusual objects at lower redshift.  Fortunately
these objects are so bright that the opportunity exists for these sources
to be measured by current near-IR spectrographs on 8-10m class telescopes
to establish their redshifts.

The $Spitzer$/IRAC flux measurements further enable the first derivation of
stellar mass estimates for $z\sim10$ galaxy candidates. The four sources
with clean IRAC photometry all show masses in the range $\log_{10} M =
8.7-9.4$ \msol\ (see Table \ref{tab:masses}). The brightest source, which is
securely detected in both IRAC bands, shows effectively no Balmer break
with its [3.6]$-$[4.5] color and overall reveals a  featureless SED,
indicative of very recent onset of star-formation.  Its UV continuum slope
is measured to be $\beta = -2.0\pm0.2$, which requires a non-negligible
amount of dust to be present already at a cosmic age of 500 Myr.   While
somewhat redder than the UV slopes reported for faint galaxies at $z\sim7$
\citep[e.g.,][]{Bouwens12a,Bouwens13,Finkelstein12}, this observed $\beta$
fits well with previously reported $\beta$-luminosity trends
\citep{Bouwens13} and is not surprising given the galaxy's luminosity.


\vspace{4pt}
\textit{The $z>8$ Luminosity Function: Luminosity or Density Evolution? }
The three highest-redshift GOODS-N sources and one of the GOODS-S sources satisfy the previously adopted
$J_{125}-H_{160}>1.2$ criterion for $z\sim10$ galaxy selections in the
GOODS-S and HUDF09/12 fields. We use these four new sources to update our
previous estimates of the $z\sim10$ UV LF and of the cosmic SFRD evolution
across $z\sim8-10$, now using the complete CANDELS-Deep dataset of GOODS-N
and South, together with the three ultra-deep HUDF09/12 fields.
The UV LF determination at $z>9.5$ thus consists of five sources: the three bright
GOODS-N candidates, one bright GOODS-S candidate, and the very faint candidate XDFj-38126243 found in the XDF.

Based on previous estimates of the UV LF at $z>8$, the detection of four
such luminous $z\sim10$ galaxy candidates in the GOODS-N+S fields is
unexpected, suggesting that the galaxy population at $z>8$ is highly
biased. 
The full $z\sim10$ sample studied here populates only
 the bright and the faint tails of the expected magnitude
distribution of an assumed Schechter function LF, with a dearth of
intermediate luminosity galaxy candidates (see, e.g., Fig \ref{fig:Nexp}).

The updated $z\sim10$ UV LF estimates are presented in Section
\ref{sec:z10UVLF}. As expected, the detection of the four new bright GOODS-N+S
sources significantly tilts the UV LF parameters to higher number densities
at bright luminosities compared to  previous work \citep{Oesch13}. The UV
LF parameters are derived relative to a baseline $z\sim8$ LF and are
summarized in Table \ref{tab:lfsummary}. In particular, we test two
scenarios for the UV LF evolution to $z\sim10$: $\phi_*$-only or
$M_*$-only. The $\phi_*$-only evolutionary scenario results in a better fit
to the current $z\sim10$ search results than the $M_*$-only. 
However, even with these new Schechter function parameters, the detection of four such bright sources is surprising given the expected number of only 1 
source at $H_{160}<27$ mag in the full search area. 
 This highlights the need for spectroscopy to determine the true nature of these bright candidates and the need to search larger areas for $z\sim9-10$ sources to determine the distribution and abundance of star-forming galaxies in the very early universe.
If these candidates are confirmed to be at high redshift, they would indicate that cosmic variance is likely larger than expected.  This could be caused, e.g., by bursty and highly biased star-formation in the very early universe \citep[see, e.g.][]{Jaacks12b,Wyithe13}.
Alternatively, this may point to a non-Schechter-like form of the LF at $z>6$ \citep[e.g.,][]{Bouwens11c,Bowler12}.

If $\phi_*$ evolution is correct and if it is not offset by a steepening in the
faint-end slope of the UV LF (i.e., $\alpha<-2$), this would result in a
significant drop in the total ionizing flux density of galaxies and may
thus have important consequences for cosmic reionization. With the small
sample in our present study, we cannot measure the evolution in $\alpha$.
However, if the LF is steeper at z$\geq$10, it would be consistent with
both the observational trends
\citep[][Bouwens et al. 2014, in prep.]{Bouwens11c,Bouwens12b,Bradley12,McLure13,Schenker13} and predictions
from theory \citep{Trenti10,Salvaterra11,Jaacks12a}.  Ultimately, larger
samples may be able to address this in the near future from the HST
Frontier Fields program.

\vspace{4pt}
\textit{The Star Formation Rate Density since $z\sim10$: }
We re-evaluated the cosmic SFRD given these new $z\sim9-10$ sources.  We
included these new sources together with all prior $z > 8$ candidates from the literature to
obtain an updated estimate of the cosmic SFRD at $z\sim10$ and found it to
be a factor $1.05\pm0.38$ dex 
 below the $z\sim8$ value. The addition of
the four luminous sources found here did not significantly change 
the rapid evolution in the SFRD that we reported in \citet{Oesch13}. This is due to the steepness of the LF and the fact that the luminosity density at $z\sim10$ (as at $z\sim6-8$) is dominated by the faint, lower
luminosity sources.  The cosmic SFRD evolves
rapidly in just 170 Myr from $z\sim10$ to $z\sim8$ (down to the current limit of $>0.7
M_\odot$~yr$^{-1}$) as shown in Figure \ref{fig:SFRDevol}.

\vspace{4pt}
\textit{The Mass Density at $z\sim10$: }
Based on the individual stellar mass measurements of the GOODS-N $z\sim10$
candidates, we attempt a first estimate of the cosmic stellar mass density
at $z\sim10$. After correcting from the bright, IRAC detected GOODS sources to LBGs
down to $M_{UV}<-18$, we find $\log_{10}\rho_{*} = 4.7^{+0.5}_{-0.8}$
\msol~Mpc$^{-3}$. This is more than an order of magnitude lower than the
current measurements at $z\sim8$. However, this decline is in good agreement with model predictions (see Fig. \ref{fig:SMDevol}).
Larger galaxy samples with direct IRAC
detections will be necessary, to verify this first stellar mass density
estimate at $z>8$ in the future.  As our analysis has shown, IRAC is
capable of detecting faint galaxies at very high redshifts. With a larger
survey it would thus be possible to significantly reduce the uncertainties
in current stellar mass density measurements at $z\geq7$.

\vspace{4pt}
\textit{Future Opportunities: }
The discovery of such luminous candidate galaxies at $z\sim10$ --
together with the existence of similarly luminous galaxies in
$z\sim6$-8 probes
\citep[][Bouwens et al. 2014, in prep.]{Bunker03,Bouwens10c,Oesch12b,Trenti11,Yan12,Bowler12} -- gives
us hope that comparably luminous sources, i.e., $-22$ to $-21$ mag,
may be found in searches at even higher redshifts, i.e., $z>10$,
albeit with substantially reduced volume densities.


If the current $z\sim9-10$ candidates are indeed at such high redshift, as we fully expect given the low likelihood of low redshift contaminants, it would argue for the use of deep but moderately wide-area searches to maximize the number of $z>8$ galaxies that are found. There are still sufficiently large uncertainties in the evolution of the UV LF at $z>6$ at present, however, that the optimal survey strategy is far from clear.   Better constraints on the UV LFs at $z\sim9$ and $z\sim10$ with present and upcoming HST data will therefore be key to inform the optimal survey strategies, e.g., for JWST.

The unusual brightness of these $z\sim9-10$ candidates makes them obvious
targets for spectroscopy, both from the ground and from space. The
brightest source is within 0.1 mag of the highly magnified CLASH $z>10$
source of \citet{Coe13}. Deep spectroscopy could reach $z>9$ Ly$\alpha$
emission lines with an equivalent width as small as 10 \AA\ and will rule
out contamination by lower level emission line sources with significant
dust extinction. Spectroscopic redshift measurements could show if these
surprisingly luminous candidates are really at high redshift as all the
photometric tests suggest. They could significantly reduce the uncertainties
on the physical parameter estimates of these sources and provide the
basis for a more detailed modeling of their spectral energy distributions
and star-formation histories. Furthermore, they would provide proof for
large variations in the number counts of very early galaxies on GOODS-size
field areas.


The bright $z\sim9-10$ candidates highlight the importance of probing a large
volume and several independent fields for accurate cosmic average
measurements at high redshift. Although somewhat reduced by lensing
magnification, the upcoming HST Frontier Fields program will provide
additional search volume for faint $z\sim10$ galaxy candidates and,
together with its $Spitzer$/IRAC component, will provide new opportunities
for exploring the cosmic frontier before the advent of JWST.

\acknowledgments{
Support for this work was provided by NASA through Hubble Fellowship grant HF-51278.01 awarded by the Space Telescope Science Institute, which is operated by the Association of Universities for Research in Astronomy, Inc., for NASA, under contract NAS 5-26555.
Additionally, this work was supported by NASA grant NAG5-7697, NASA grant
HST-GO-11563, NASA grant HST-GO-12177, NASA grant JPL-1416188/1438944, ERC
grant HIGHZ \#227749, and NWO vrij competitie grant 600.065.140.11N211. 
This work was further supported in part by the National Science Foundation under
Grant PHY-1066293 and the Aspen Center for Physics. 
Some of the data presented in this paper were obtained from the Mikulski Archive for Space Telescopes (MAST). This research used the facilities of the Canadian Astronomy Data Centre operated by the National Research Council of Canada with the support of the Canadian Space Agency. 
}

Facilities: \facility{HST(ACS/WFC3), Spitzer(IRAC)}.

\bibliography{MasterBiblio}
\bibliographystyle{apj}



%

\appendix

\section{Two Bright $z>9$ Candidate Galaxies in GOODS-S}

Motivated by the discovery of the four bright $z>9$ galaxy candidates in the GOODS-N field, we systematically re-analyzed the GOODS-S dataset. In particular, we ran new SExtractor catalogs with a higher deblending efficiency to split neighboring sources, and we searched for additional sources with a bluer color cut of $J_{125}-H_{160}>0.5$, as we did in
GOODS-N, rather than the more conservative cut of $J_{125}-H_{160}>1.2$ as adopted in our previous work \citep[e.g.][]{Oesch13}.

These new catalogs revealed two possible, bright $z>9$ galaxy candidates in the CANDELS GOODS-S dataset, GS-z9-1 and GS-z10-1. They have magnitudes of $H_{160} = 26.6\pm0.2$ and $H_{160}=26.9\pm0.2$, respectively. The latter candidate also shows a color of $J_{125}-H_{160}>1.2$ (namely $1.7\pm0.6$), while the first is only slightly too blue to satisfy this criterion ($J_{125}-H_{160}=1.1\pm0.5$).

Given its red color, GS-z10-1 could already have been in the previous catalog of \citet{Oesch13} who analyzed the same CANDELS GOODS-South dataset. The reason this source was not previously selected is due to a very faint neighbor that was included in the Kron aperture in the earlier SExtractor catalog. This caused the candidate to be rejected due to apparent optical flux in the aperture. With careful visual inspection we assessed that the optical flux in the previous aperture was due to a faint neighboring galaxy and is not likely associated with the high-z candidate. With the new deblending parameters for our SExtractor run, this source is now confirmed to be a legitimate $z>9$ galaxy candidate. Its photometric redshift is found to be $z_\mathrm{phot} = 9.9\pm0.5$. We thus include this candidate in the full analysis of the main body of this paper, and we also updated the completeness and selection functions corresponding to our new SExtractor catalogs. 

The inclusion of this $z\sim10$ candidate does not significantly change the results. For instance, including this candidate only causes a change of $0.1$ dex in $\phi*$ when assuming density evolution or a change of only 0.1 in $M*$ for luminosity evolution. The total cosmic SFRD changes by only 0.02 dex, because this is dominated by large flux from lower luminosity sources as indicated by the faintest candidate in the XDF (and by the steep slopes
found at slightly later times at $z\sim7-8$).

The other source, GS-z9-1, was already in the previous SExtractor catalogs. However, it was not included in the analysis due to its bluer color of $J_{125}-H_{160}<1.2$. For completeness, we present this source here as well, particularly since it is so close to our $z\sim10$ color cutoff. Interestingly, it also shows significant IRAC detections in both 3.6 and 4.5~$\mu$m bands with fluxes consistent with a significant Balmer break at $z\sim9$, giving added weight to our identification of this source as a probable $z\sim9$ candidate. From SED fitting we find a photometric redshift of $z_\mathrm{phot} = 9.3\pm0.5$ for this source. 

\begin{figure*}[htp]
	\centering
	\includegraphics[width=0.7\linewidth]{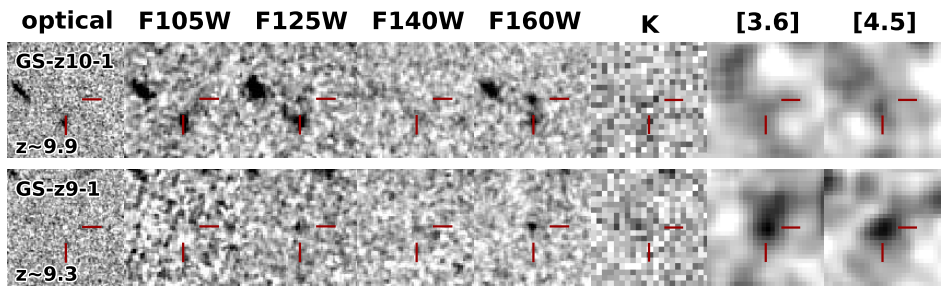}
  \caption{6\arcsec$\times$6\arcsec\ negative images of the two new $z\geq9$ galaxy candidates identified in our re-analysis of the CANDELS GOODS-S data. From left to right, the images show a stack of all optical bands, \yFilter, \jFilter, \hFilter, HAWKI K, and neighbor-subtracted IRAC 3.6\,$\mu$m and 4.5\,$\mu$m images. The K-band image is a very deep stack (26.5 mag, 5$\sigma$) of ESO/VLT HAWK-I data from the HUGS survey (PI: Fontana). Both sources are only weakly detected in these data. }
	\label{fig:stamps_newcandidates}
\end{figure*}

\begin{figure}[tbp]
	\centering
\includegraphics[width=0.6\linewidth]{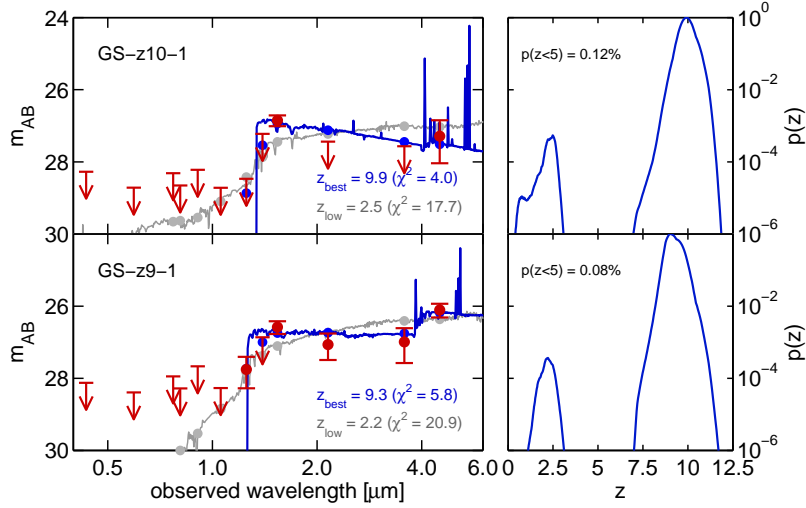}	  
  \caption{Spectral energy distribution fits (left) and redshift likelihood functions (right) for the two bright $z\sim9-10$ sources in GOODS-S. 
  In the left panel, photometry and $2\sigma$ upper limits are shown in dark red. 
Best-fit SEDs are shown as blue solid lines, in addition to the best low redshift solutions in gray. The legend lists their respective redshift and $\chi^2$ values. The redshift likelihood distributions from SED fitting in the right panels are shown on logarithmic axes.  For both sources, the likelihood of a low redshift solution is only $\sim1\%$ as indicated by the labels.  }	
\label{fig:sedFits_newcandidates}
\end{figure}

Images of both new GOODS-S candidates are shown in Figure \ref{fig:stamps_newcandidates}, and their SED fits and photometric redshift likelihood functions are shown in Figure \ref{fig:sedFits_newcandidates}. Table \ref{tab:newcandidatesGS} lists the basic information of these sources, and Table \ref{tab:fluxNewCand} list all their flux measurements.

\begin{deluxetable*}{llcccccc}
\tablecaption{Coordinates and Basic Photometry of two new $z>9$ LBG Candidates in the GOODS-S Field\label{tab:newcandidatesGS}}
\tablewidth{\linewidth}
\tablecolumns{8}

\tablehead{\colhead{Name} & \colhead{ID} & RA & DEC &\colhead{$H_{160}$}  &\colhead{$J_{125}-H_{160}$}  & \colhead{$H_{160}-[4.5]$}   &  \colhead{$z_{phot}$} }

\startdata

GS-z10-1 & GSDJ-2269746283 &  03:32:26.97 & -27:46:28.3  & $26.88\pm0.15$  & $1.7\pm0.6$ & $-0.4\pm0.6$ & $9.9\pm0.5$  \\

\noalign{\vskip .7ex} \hline \noalign{\vskip 1ex}
GS-z9-1 & GSDJ-2320550417* & 03:32:32.05 & -27:50:41.7  & $26.61\pm0.18$  & $1.1\pm0.5$ & $0.5\pm0.3$ & $9.3\pm0.5$  


\enddata

\tablenotetext{*}{The source GS-z9-1 does not satisfy the criterion $J_{125}-H_{160}>1.2$ and is not included in the UV LF analysis.}
%
\end{deluxetable*}

\begin{deluxetable}{lcc}
\tablecaption{Flux Densities of Two New $z>9$ LBG Candidates in the GOODS-S Field\label{tab:fluxNewCand}}
\tablewidth{\linewidth}
\tablecolumns{3}

\tablehead{Filter  & GS-z10-1 &GS-z9-1  } 
\startdata

$B_{435}$  & $-1\pm9$  &  $7\pm10$ \\ 
$V_{606}$  & $1\pm6$  &  $0\pm8$ \\ 
$i_{775}$  & $-6\pm9$  &  $-5\pm12$ \\ 
$I_{814}$  & $5\pm6$  &  $-3\pm9$ \\ 
$z_{850}$  & $-4\pm9$  &  $-5\pm16$ \\ 
$Y_{105}$  & $0\pm6$  &  $-14\pm9$ \\ 
$J_{125}$  & $13\pm7$  &  $29\pm11$ \\ 
$JH_{140}$  & $12\pm23$  &  $55\pm33$ \\ 
$H_{160}$  & $66\pm9$  &  $85\pm14$ \\ 
$K\mathrm{-HAWKI}$  & $33\pm19$  &  $54\pm18$ \\ 
 IRAC 3.6\,$\mu$m & $32\pm17$  &  $58\pm24$ \\ 
 IRAC 4.5\,$\mu$m  & $44\pm22$  &  $131\pm23$
 

 \enddata

 \tablecomments{Measurements are given in nJy with $1\sigma$ uncertainties.}
 
 
\end{deluxetable}

\end{document}